\documentclass[preprint,12pt]{elsarticle}

\usepackage{amsmath,amssymb}
\usepackage{graphicx}
\usepackage{xcolor,setspace,relsize,mathrsfs,xstring,float,hyperref,xparse,ifthen,caption}
\usepackage{subcaption}
\usepackage{placeins}
\usepackage{rotating}
\usepackage{dashrule}
\usepackage{lineno}
\usepackage{microtype}

\definecolor{red}{rgb}{1,0,0}
\definecolor{pink}{rgb}{1,0,1}
\definecolor{dgreen}{rgb}{0,0.6,0}
\definecolor{green}{rgb}{0,1,0}
\definecolor{blue}{rgb}{0,0,1}
\definecolor{dblue}{rgb}{0,0,0.6}
\definecolor{magenta}{rgb}{1,0,1}
\definecolor{cyan}{rgb}{0,1,1}
\definecolor{pakistangreen}{rgb}{0.0, 0.4, 0.0}

\newcount\ndots
\def\drwln#1#2{\raise 2.5pt\vbox{\hrule width #1pt height #2pt}}
\def\spc#1{\hskip #1pt}
\def\solid{\drwln{24}{1.0}\ }
\def\chndot{\hbox {\drwln{5}{1.0}\spc{2}\drwln{2}{1.0}\spc{2}\drwln{5}{1.0}}\nobreak\ }
\def\dashed{\hbox {\drwln{4}{1.0}\spc{2}\drwln{4}{1.0}\spc{2}\drwln{4}{1.0}}\nobreak\ }
\def\dotted{\hbox {\drwln{2}{1.0}\spc{2}\drwln{2}{1.0}\spc{2}\drwln{2}{1.0}}\nobreak\ }

\newcommand{\dpr}[2] {\frac{\partial{#1}}{\partial {#2}}}

\DeclareSymbolFont{AMSa}{U}{msa}{m}{n}
\let\square\undefined
\DeclareMathSymbol{\square}{\mathord}{AMSa}{"03}

\newcommand{\be}{\begin{equation}}
\newcommand{\ee}{\end{equation}}
\newcommand{\beq}{\begin{equation}}
\newcommand{\eeq}{\end{equation}}
\newcommand{\bea}{\begin{eqnarray}}
\newcommand{\eea}{\end{eqnarray}}

\journal{Journal of Fluids and Structures}

\begin{document}

\sloppy
\raggedbottom

\begin{frontmatter}

\title{Modeling and Computational Fluid Dynamics Validation of a Nonholonomically Constrained Two-Rigid-Body Swimming System}

\author[msu]{J. Ardister}
\author[msu]{J. Geddes}
\author[msu]{B. F. Feeny}
\author[msu]{J. Yuan\corref{cor1}}
\ead{junlin@msu.edu}

\affiliation[msu]{organization={Department of Mechanical Engineering, Michigan State University},
            addressline={428 S Shaw Ln},
            city={East Lansing},
            postcode={48824}, 
            state={MI},
            country={USA}}

\cortext[cor1]{Corresponding author}

\begin{abstract}
A simple nonholonomic dynamics model is developed as a low-order model for generating undulatory swim-like motions, validated through computational fluid dynamics (CFD) simulations. The rigid-body-dynamics model generates swimming motion by imposing a nonholonomic (NH) constraint on the tail of a two-body system, requiring that tail-fin velocity aligns with the tail angle, while the head moves in a straight line through a slot constraint. The system has one degree of freedom, with equations of motion derived using Lagrange multipliers. Two-dimensional CFD simulations validate the model in an incompressible Newtonian fluid, where the resolved tail fin interacts with fluid through the immersed boundary method until steady-state swimming is achieved. The validation demonstrates excellent quantitative agreement between CFD and model predictions for body orientation angle and normal fluid force across variations in fin motion amplitude, period, and Reynolds number. While an exact NH constraint point does not exist, an effective period-averaged NH location can be identified for successful model predictions. At higher Reynolds numbers, the two-body kinematics displays independence from the Reynolds number variation. The CFD data reveal that the two-body model captures the type of power-law relationship between Reynolds and Strouhal numbers governing undulatory swimming from tadpoles to whales, indicating that the simplified two-link model is representative of swimming dynamics in continuous geometries at various scales. A key limitation is that the drag force model requires \textit{a priori} CFD calibration to match steady-swim velocity, limiting standalone predictive capability. The results demonstrate that the low-order NH constraint-based model effectively captures essential swimming dynamics, offering a robust alternative to existing fluid-force models.

\end{abstract}


\end{frontmatter}

\section{Introduction}

The study of aquatic locomotion has gained significant attention in the context of biomimicry and efficient swimming mechanisms. Carangiform swimmers \cite{Lindsey78,Webb78,Videler93}, with their undulatory motion and flexible bodies, provide an insightful case for examining the relationship between fluid dynamics and body mechanics. Breder \cite{breder1926} and Taylor \cite{Taylor1952} established key principles of fish locomotion. Gray made detailed observations of swimming patterns, laying the groundwork for many kinematic models \cite{Gray1933a,Gray1933b}. Videler and Hess \cite{VidelerHess1984} and Lauder \cite{lauder2007} have computationally analyzed the hydrodynamics of efficient propulsion, revealing critical insights into optimizing body shape and fin morphology.

Understanding fish swimming requires modeling the interaction between body mechanics and fluid dynamics. Low-order models typically focus on the influence of the fluid on the structure, using idealized force representations. The Taylor model \cite{Taylor1952}, for instance, is a resistive force model based on empirical measurements of steady flow over slender bodies. It applies viscous drag forces to a moving fish under the assumption that unsteady effects are negligible. Variants of this model have been used in lumped-mass mechanical representations of fish \cite{McMillenHolmes2006,McMillenMWH2008,BhallaBGP13}. In contrast, Lighthill's model \cite{Lighthill60} describes the reactive forces generated by body motion, assuming inviscid flow while emphasizing added mass effects associated with unsteady fluid displacement.  This model has been examined, for example, through lamprey simulations by Tytell et al. \cite{TytellTHWCF10} and employed in optimization frameworks by Eloy \cite{EloyE13} to study swimming efficiency.

Despite their utility, these traditional models face several fundamental limitations that restrict their predictive accuracy and applicability. Resistive force models \cite{Taylor1952} suffer from quasi-steady assumptions that neglect fluid inertia and time-history effects. Lighthill's reactive force theory \cite{LighthillL71}, while incorporating added-mass effects, relies on small-amplitude and slenderness approximations that break down for large-amplitude tail motions and cannot account for vortex shedding phenomena \cite{LighthillL71}. The large-amplitude extension by Lighthill \cite{LighthillL71} and subsequent modifications by Wu \cite{Wu61} and Candelier et al. \cite{CandelierCBL11}, among others, improved certain aspects but maintained the inviscid flow assumption.
Recent works have addressed some of these limitations, with some examples as follows. Eloy \cite{EloyE13} developed a combined reactive-resistive model that was used to identify optimal swimming designs of fish; Tytell et al. \cite{TytellTHWCF10} used fully coupled simulations to quantitatively test classical theories, revealing significant discrepancies in force predictions; and high-fidelity CFD studies by Borazjani and Sotiropoulos \cite{borazjani2008,BorazjaniSotiropoulos09} demonstrated the importance of three-dimensional vortical flows missing from traditional models.

Nonholonomic (NH) constraint-based models provide an alternative simplification of fin effects. NH constraints impose velocity restrictions on body motion (such as preventing lateral slipping at a fin) without directly specifying positional relationships. This approach models fins as ``keels" cutting through water, much like the keel of a sailboat, leading to constraints that cannot be integrated into coordinate-only expressions. NH constraints have been applied in fish-like swimming models \cite{CochranEtal09,
BazziEtal2017,PollardFT19,ArdisterFeeny2022a,ArdisterFeeny2022b,ArdisterFeeny2023}. These studies highlight the potential of NH constraints to capture essential aspects of swimming dynamics while avoiding the computational expense of full fluid-structure interaction (FSI) simulations. The key distinction between NH constraints and resistive/reactive force models lies in their physical modeling assumptions: NH constraints apply constraint-based mechanics, while resistive and reactive models rely on formulations of forces and momentum under specific flow regime assumptions. This distinction enables the NH constraint approach to potentially offer broader applicability across flow conditions as long as the constraint holds. Compared to prior NH constraint studies, the present work focuses on quantitative validation against high-fidelity CFD simulations.

Here, we investigate the validity of the NH constraint by studying a two-rigid-body ``slot-car" model of fish-like locomotion consisting of a ``head" and a ``tail" rigid body. The mid-point of the head body is constrained to move along a straight line (analogous to a slot-car track), while the tail oscillates relative to the head and interacts with the fluid through the NH constraint at the fin location. A single NH constraint is imposed at the tail fin, providing a simple mechanism that generates effective locomotion. This configuration can be viewed as a minimal multi-body dynamics model for fish-like locomotion: a two-link chain with a single mechanical degree of freedom, in contrast to the standard multi-degree-of-freedom chains used in existing swimming studies \citep[][among others]{BhallaBGP13,EkebergGrillner99}. In these studies, the fish-like swimmers are approximated as chains of rigid segments with multiple joints and internal actuation, often coupled to simplified hydrodynamic models, CFD or experiments. We deliberately adopt this minimal configuration as a validation-first approach to enable clean assessment of the NH constraint mechanism: if the framework can accurately accommodate fluid-structure interactions in this simplest realization without confounding factors from body complexity or multiple fin interactions, it justifies subsequent application to more complex multi-link formulations.

We derive the strongly nonlinear equations of motion using Lagrange multipliers, approximate them to cubic nonlinearity, and analyze the system's behavior using harmonic balance method and numerical solutions. We then conduct CFD simulations to validate the NH constraint as a low-order model of fin-fluid interaction in this simple mechanism, using parameters matched to the low-order mechanics model as closely as possible while acknowledging necessary concessions. This comparison allows us to examine how Reynolds number, tail-fin oscillation amplitude and oscillation frequency affect swimming dynamics, as  understanding parameter sensitivity in swimming dynamics is important \cite{McMillenHolmes2006,kelly2009}.

This work has two primary aims: (1) to demonstrate that a simple model with two rigid bodies and a single NH constraint can generate swimming-like locomotion patterns,  and (2) to evaluate the validity of modeling assumptions by comparing results to CFD simulations that capture full fluid-structure interaction.
If the NH constraint is validated as a reasonable approximation of fluid-structure interaction in this idealized setting, we hypothesize that this concept can be extended to more realistic models of fish locomotion without slot constraints and with multiple fins. 

In what follows, we derive the nonholonomic low-order model of undulatory locomotion with a slot constraint, analyze its dynamics, characterize its behavior in effective swimming regimes, and present CFD-based evaluations of its assumptions. A deeper understanding of low-order fin-fluid models will have significant implications for biological studies and biomimetic design, and potentially other applications \cite{LiuQuSongChen2025,MylonasSayer2012,WangEtal2023,KimEtal2010}.

The remainder of this paper is organized as follows. Section 2 derives the equations of motion for the nonholonomic two-body model using Lagrangian mechanics and presents analytical approximations. Section 3 characterizes the model's behavior and examines parameter effects, demonstrating the system's ability to generate swimming-like locomotion patterns (Aim 1). Section 4 presents CFD simulations to validate the nonholonomic constraint as an effective model of fin-fluid interaction (Aim 2), comparing predictions with full fluid-structure interaction results. Section 5 further supports Aim 1 by showing that the model reproduces the fundamental Reynolds-Strouhal scaling relationship observed across biological swimmers. Section 6 discusses key insights, limitations including the need for \textit{a priori} drag calibration, and broader implications. Section 7 concludes the manuscript.

\section{Low Order Modeling}
\newcommand{\figwidth}{0.9\linewidth}

\subsection{Equations of Motion}
\label{sec:model_eq}

\begin{figure}[t]
    \centering
    \includegraphics[width=\figwidth]{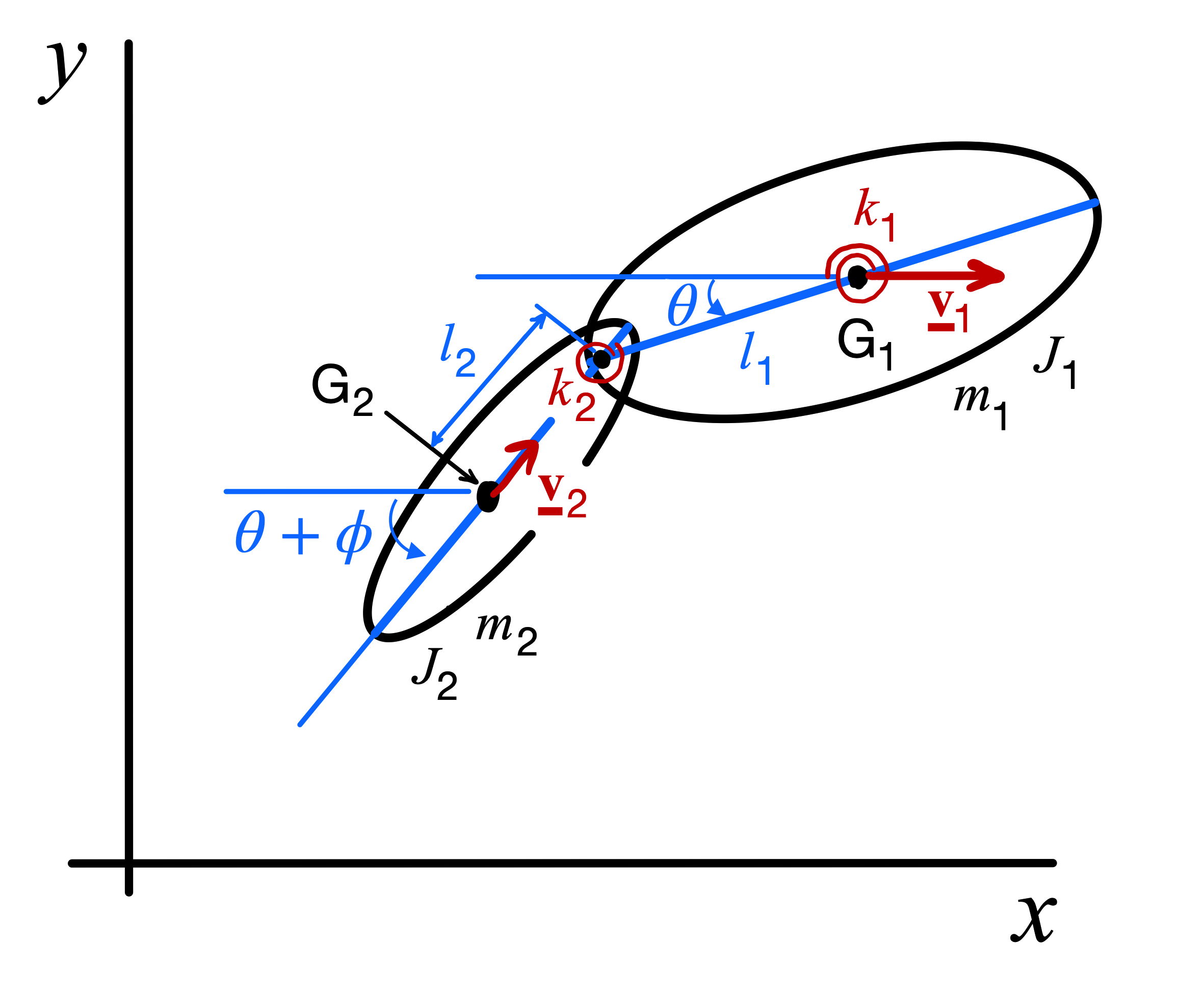}
    \caption{Schematic of the two-rigid-body fish approximation, illustrating the coordinate system, key parameters, and constraints. The head (Body 1) and tail (Body 2) are connected by a frictionless link, with the tail's velocity constrained nonholonomically to align with its absolute angle, $\theta + \phi$.}
    \label{fig:schematic}
\end{figure}

We model a system of two rigid bodies connected by a frictionless link, as illustrated in Figure~\ref{fig:schematic}. Body 1, or the ``head," has a total length of $2l_1$, a total mass of $m_1$, and a mass moment of inertia $J_1$ about its mass center, $G_1$. The position $(x_1, y_1)$ of $G_1$ is constrained such that $y_1 = 0$, a condition referred to as the \textit{slot constraint}, allowing translation along the $x$-axis and rotation about $G_1$. The head's absolute angle, $\theta$, is measured from the horizontal axis.

Body 2, or the ``tail," has a total length of $2l_2$, a total mass of $m_2$, and a mass moment of inertia $J_2$ about its mass center, $G_2$. The angle of the tail relative to the head is $\phi$, such that its absolute rotation is given by $\theta + \phi$. The velocity of $G_2$, denoted by $\vec{v}_2$, is subject to a nonholonomic constraint, so that its direction aligns with the absolute angle of the tail, $\theta + \phi$.

The model incorporates the following assumptions:
\begin{enumerate}
    \item \textbf{Zero angle of attack:} The tail fin operates at approximately zero angle of attack: the angle between the fin orientation and the fin's relative velocity to the freestream  is negligible.
    \item \textbf{Point fin only:} The tail fin is approximated as a point (i.e., $G_2$), and its interaction is modeled as a constraint in which the fin moves in the direction in which Body 2 is aligned (applying the first assumption).
    Thrust forces from the 
    rest of the swimming body are neglected.
    \item \textbf{Mass:} The bodies have uniform mass distribution, with their centers of mass at their geometric centers.
    \item \textbf{Drag Model:} 
    A fluid drag force is applied to the head body and only resists forward motion. For simplicity, we assume that fluid drag predominantly arises from the fin and use the laminar flat-plate skin-friction law (Blasius solution)  \cite{schlichting1979} to estimate the frictional drag on the fin. 
\end{enumerate}

The slot constraint is devised to keep the body moving straight under the action of a single fin, providing the simplest model of fluid-structure interaction for validation with CFD. 
Due to the complexity introduced by nonholonomic constraints, which are generally nonlinear in nature, we employ specialized approaches to derive equations of motion for such systems. These include nonholonomic Lagrange equations \cite{Feeny06a}, Lagrange multipliers, the principle of virtual power \cite{Moon2008}, and Newton's method, which are used to verify the derived equations.

In applying Lagrange multipliers \cite{Greenwood1988}, we initially relax the nonholonomic constraint while enforcing the holonomic (geometric) constraints. The positions of the centers of mass $G_1$ and $G_2$ are given as
$\vec{r}_1=\left[x_1,y_1\right]$ and $\vec{r}_2=\left[x_2,y_2\right]$,
and the bodies 1 and 2 have rotations $\theta$ and $\theta+\phi$. The slot constraint is $y_1 = 0$, and the link constraints are
\begin{align}
    \label{eq:r2xy}
    x_2 & = x_{1} - l_{1}\cos{\left(\theta\right)} - l_{2}\cos{\left(\theta+\phi\right)}, \nonumber \\
    y_2 & = y_{1} - l_{1}\sin{\left(\theta\right)} - l_{2}  \sin{\left(\theta+\phi\right)}.
\end{align}
The NH constraint on the tail is such that $\dot y / \dot x = \tan(\theta+\phi)$, which can be expressed as
\begin{equation}
    \label{eq:NHConstraint1}
    \dot{x}_2 \sin(\theta + \phi) - \dot{y}_2 \cos(\theta + \phi) = 0.
\end{equation}
Differentiating Eq.~\eqref{eq:r2xy}, we can write $\dot y_2$ in terms of $\dot \theta$. In this way, generally $p$ constraint equations can be written as
\begin{equation} \label{eq:constraintGeneralForm}
    \sum_{k=1}^{n+p} a_{hk} \dot q_k + a_{h0} = 0, \qquad h = 1, \dots, p.
\end{equation}
where coefficients $a_{hk}$ and $a_{h0}$ are functions of the coordinates and time, and the $q_k$ are generalized coordinates remaining after eliminating holonomically constrained coordinates. In this problem, $q_1 = x_1$, $q_2 = \theta$, and $p=1$.  We can solve Eq.~\eqref{eq:constraintGeneralForm} for $\dot \theta$ and reduce the NH constraint to the form
\begin{equation} \label{eq:NHconstraintForm}
    \dot \theta = g(\dot x_1, \theta, \phi, \dot \phi, \ddot \phi),
\end{equation}
by which we have chosen $\theta$ (through $\dot \theta$) to be a dependent coordinate. Finally, in this model, we impose the relative angular displacement, such that $\phi = a \sin \omega t$, where $a$ is the amplitude and $\omega$ is the angular frequency of tail oscillation, thereby constraining $\phi$.
In summary, among $(x_1, y_1, x_2, y_2, \theta, \phi)$, with constraint equations $y_1 = 0$, $a \sin \omega t$, Eq. \eqref{eq:r2xy}, and Eq.\eqref{eq:NHconstraintForm}, we have $x_2 = x$ remaining as the independent generalized coordinate. This means our two-rigid-body approximation of undulatory fish-like swimming has a single degree of freedom, i.e. $n=1$.

The Lagrangian is defined as $\mathscr{L} = T - V$, where $T$ is the total kinetic energy and $V$ is the total potential energy. To use the Lagrange multiplier formulation, we relax specified constraints and then formulate the ``unconstrained'' Lagrangian. For this system, we relax only the NH constraint, yielding
\begin{equation}
    \label{eq:lagrangian1}
    \mathscr{L}^u = \frac{1}{2} \sum_{i=1}^n \left[ m_i \left( v_i^{\;u} \right)^2 + J_i \left( \dot{\theta}_i^{\;u} \right)^2
        - k_i \left( \phi_i^{\;u} \right)^2 \right].
\end{equation}
In this system, $N=2$, $i$ indexes bodies 1 and 2, $m_i, J_i, k_i, \dot\theta_i,$ and $\phi_i$ represent mass, mass moment of inertia, stiffness if present, absolute angular velocity, and relative angular displacement, of body $i$.  $v_i^u = |\dot{\vec r}_i^{~u}|$ is the velocity of the COMs of body $i$ after holonomic constraints have been substituted, and is therefore a function of $x, \dot x, \theta, \dot\theta$, and $\phi(t)$.  The superscript $u$ indicates that the quantity should be ``unconstrained" with respect to the relaxed nonholonomic constraint.

In addition to the Lagrangian, we also need to formulate the generalized forces of the system. To achieve this, we will use virtual nonconservative work, $\delta W^{nc}$, and unconstrained virtual displacements, $\delta \vec{r}^{~u}_i$ \cite{goldstein2001}. The unconstrained virtual displacement vectors relax specified constraints (in our case the NH constraint) and can be written as
\begin{equation}
    \label{eq:VirtualDisplacement}
    \delta\vec{r}^{\;u}_{i}=\sum_{k=1}^{n+h}\left[\frac{\partial\vec{r}^{\;u}_i}{\partial q_k}\delta q_k\right],
\end{equation}
where $q_k$ are the unconstrained generalized coordinates and $\vec{r}^{~u}_i$ are the position vectors described in the setup section. The virtual work is
\begin{equation}
    \label{eq:VirtualWork1}
    \delta W_{nc}=\sum_{i=1}^{N}\left[{\vec{F}_i}^{nc}\cdot\delta \vec{r}_i+\vec{M}_i\delta\theta_i\right],
\end{equation}
where ${\vec{F}_i}^{nc}$ are the resultant nonconservative forces, due to damping or drag on body $i$, and $\vec{M}_i$ are moments on the body. With some substitution and massaging, we can represent the virtual work of the system in terms of the generalized coordinates as
\begin{equation}
    \label{eq:VirtualWork2}
    \delta W_{nc}=\sum_{k=1}^{n+h}Q_k\delta q_k,
\end{equation}
where $Q_k$ are the generalized forces used to complete the Lagrange equations.
The primary contribution to $Q_k$ will be the fluid drag applied to the head. 

We model the fluid drag according to the Blasius' laminar friction drag law  \cite{schlichting1979}: $ D = C_D \rho_f A_f |\dot{x}|^{2}$ with $C_D=a \text{Re}_L^{-1/2}$, where  $\rho_f$ is the fluid density and $a$ is a geometry-dependent constant calibrated from CFD based on the hydrofoil fin geometry (see Sec.~\ref{sec:simulation-methodologies}) and $A_f$ is the fin planform area. As such, $D\sim |\dot{x}|^{3/2}$. This type of model is widely used as a baseline viscous drag model for streamlined swimming bodies in the literature.
for example, Eloy \cite{EloyE13} assumed  laminar boundary layers along the fish body and employed an elongated-body approximation via Mangler’s transformation of the flat-plate solution. Toki\'{c} and Yue \cite{TokicY12} used a flat-plate skin-friction coefficient (corrected for thickness effects) with a Blasius-type scaling and accounted for transition from laminar to turbulent flow. The experiments of Anderson et al. \cite{AndersonAMG01} provided justifications for models that assume laminar-like friction: many fish operate in a regime where either the boundary layer is laminar or the fish’s kinematics help keep it closer to laminar flow than it would be otherwise.

To obtain the equations of motion, we use Lagrange's equations with Lagrange multipliers \cite{goldstein2001,Greenwood1988,Moon2008}, which are associated with constraint forces (or moments) that do no work,
\begin{align}
    \label{eq:LM}
    \frac{d}{dt}\left( \frac{\partial \mathscr{L}^{u}}{\partial \dot{q}_k} \right) - \frac{\partial \mathscr{L}^{u}}{\partial q_k} & =
    \sum_{h=1}^{p} \left[ \lambda_h \; a_{hk} \right] + Q^u_k,
\end{align}
\noindent for $k=1, 2$, with $q_1 = x$ and $q_2 = \theta$, while other coordinates were eliminated using the holonomic constraint, and where $p$ is the number of relaxed constraints (here $p=1$ is NH constraint).  
$\lambda_h$ is the Lagrange multiplier associated with the $h^{th}$ constraint, and $a_{hk}$ are constraint coefficients obtained from Eq. \eqref{eq:constraintGeneralForm}.

The $q_2 = \theta$ equation represents a moment balance about $G_1$, and it turns out that $a_{1\theta}$ is the moment arm from $G_1$ to the constraint point.
Hence, $\lambda_h$ (or $\lambda$ for simplicity as $p=1$) specifically reflects the constraint force at the constraint point and normal to the tail body.
Physically, the constraint force $\lambda$ is a surrogate for the resultant normal force $F_n$ of the pressure distributed over the surface of the tail of the swimmer, normal on average to the surface, and acting through its center of pressure. The $x$-component of $\lambda$ for the NH model (and $F_n$ for the fluid swimmer) provide thrust.
Analyzing this force will enable us to compare fluid forces acting on the tail fin of the fish.

For this system, this process will produce one independent second-order ordinary differential equation from Eq.~\eqref{eq:LM}, along with the differential constraint equation \ref{eq:NHconstraintForm}, as
\begin{equation}
    \label{eq:accel1}
    \ddot{x}=f\left(\dot{x},\theta,\phi(t),\dot{\phi}(t),\ddot{\phi}(t)\right),
\end{equation}
\begin{equation}
    \label{eq:ThetaDot3}
    \dot{\theta }=g\left(\dot{x},\theta,\phi(t),\dot{\phi}(t),\ddot{\phi}(t)\right),
\end{equation}
and one algebraic equation representing the constraint force
\begin{equation}
    \label{eq:constraintforce1}
    \lambda=\lambda\left(\dot{x},\theta,\phi(t),\dot{\phi}(t),\ddot{\phi}(t)\right).
\end{equation}
Equations of motion \eqref{eq:accel1} and \eqref{eq:ThetaDot3}, not displayed for brevity, were confirmed with independent derivations using the principle of virtual power \cite{Moon2008} and the nonholonomic Lagrange equations \cite{Feeny06a}.
Thus, this single-degree-of-freedom is represented by a third-order set of equations.  Letting $v=\dot x$, and with imposed $\phi(t)$, we have a three-dimensional extended state space.

\subsection{Analysis}

To make the equations amenable to analysis, we expand the nonlinear terms contained in functions $f$ and $g$ in equations \eqref{eq:accel1} and \eqref{eq:ThetaDot3} for small angles and amplitudes, and we approximate them as polynomials up to cubic degree, as the minimum nonlinearity required to capture the important dynamics \cite{nayfeh2000}.  In fact, the approximation of the equation of motion \eqref{eq:accel1} for $\ddot x$ is quadratic, and equation \eqref{eq:ThetaDot3} for $\dot \theta$ is cubic.  Both equations also have parametric terms.

Staging a harmonic-balance analysis for fluctuations, we assert an approximate steady-state solution of the form:
\begin{equation}
    \label{eq:hb_v}
    \dot{x}\left(t\right)=v=U_s,
\end{equation}
\begin{equation}
    \label{eq:hb_theta}
    \theta\left(t\right)=\theta_{c}\cos(\omega t)+\theta_{s}\sin(\omega t),
\end{equation}
where $U_s$ is the mean steady-state velocity, $\theta_{c}$ and $\theta_{s}$ are the amplitudes of the first harmonic for the response of $\theta$, and $\omega$ is the frequency of the imposed tail motion.

The following section evaluates how well the slot-car model generates forward locomotion and constraint-driven propulsion across a range of input amplitudes and frequencies. By comparing the full ODE simulation to the harmonic balance approximation, we assess the validity of truncating the system response to a single harmonic. We also quantify the conditions under which this reduced-order representation remains accurate.

\section{Behavior and Parameter Effects}

To evaluate the behavior of the slot-car model, we simulate its dynamics across a range of input amplitudes and frequencies. The primary goal is twofold: (1) to confirm that the model generates forward locomotion with realistic kinematics consistent with biological swimming, and (2) to assess the accuracy of the harmonic balance approximation through comparison with simulations of the full ordinary differential equations (ODEs) of Eqs. \eqref{eq:accel1} and \eqref{eq:ThetaDot3}.  Additionally, the behavior of the slot-car model will be used to evaluate its suitability as a reduced-order representation of undulatory locomotion in fluid, by quantifying how well the simplified solutions reproduce key dynamic features—such as velocity, orientation, and thrust generation.

\subsection{Behavior of Baseline Model}

\begin{figure}[htbp]
    \centering
    \includegraphics[width=\figwidth]{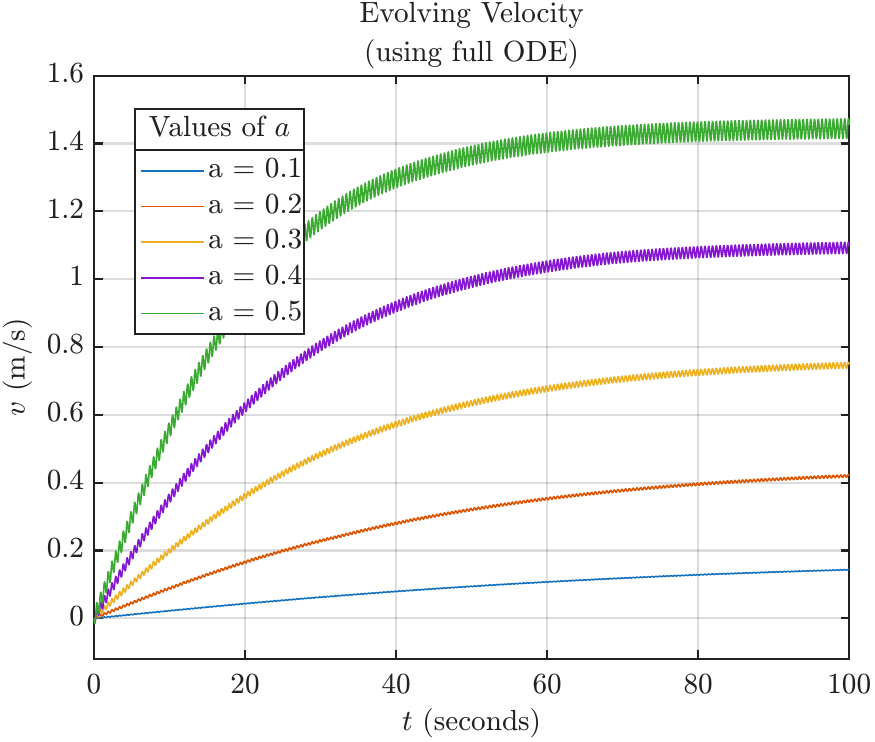}
    \caption{Time evolution of forward velocity for the slot-car model using parameter group 1. Each curve corresponds to a different input amplitude $a$, illustrating the effect of tail oscillation strength on steady-state propulsion. The system starts from rest and converges to a positive mean velocity with small superimposed oscillations.}
    \label{fig:varied_a_xdot_full}
\end{figure}

Using the parameters in Table~\ref{tab:parameter_group1}, we simulate the model starting from rest. As shown in Figure~\ref{fig:varied_a_xdot_full}, the system achieves a positive steady-state velocity with small oscillations, confirming net forward propulsion. The plot includes curves for multiple input amplitudes $a$, illustrating how increased tail motion results in stronger propulsion and larger steady-state oscillation amplitude in this range of parameters. The mean velocity magnitudes fall within a physically realistic range for a swimmer of this scale.

\begin{table}[htbp]
    \centering
    \caption{Parameter values for the two-rigid-body fish model (Figure~\ref{fig:schematic}) used in the following analysis. These values are matched dimensionlessly in the computational fluid dynamics simulations in Sec.~\ref{sec:cfd}.}

    \label{tab:parameter_group1}
    \begin{tabular}{|c|c|c|}
        \hline
        \textbf{Parameter}              & \textbf{Body 1 (head)}   & \textbf{Body 2 (tail)}        \\
        \hline
        Mass (kg)                       & 5.0                      & 0                             \\
        Length (m)                      & 0.667                    & 0.333                         \\
        Moment of Inertia (kg$\cdot$m²) & 1.25                     & 0                             \\
        \hline
        \textbf{Excitation}             & \textbf{Amplitude ($a$)} & \textbf{Frequency ($\omega$)} \\
        \hline
        $\phi(t)=a\sin(\omega t)$       & 0.2                      & 2 $\pi$                        \\
        \hline
    \end{tabular}
\end{table}

\begin{figure}[htbp]
    \centering
    \includegraphics[width=\figwidth]{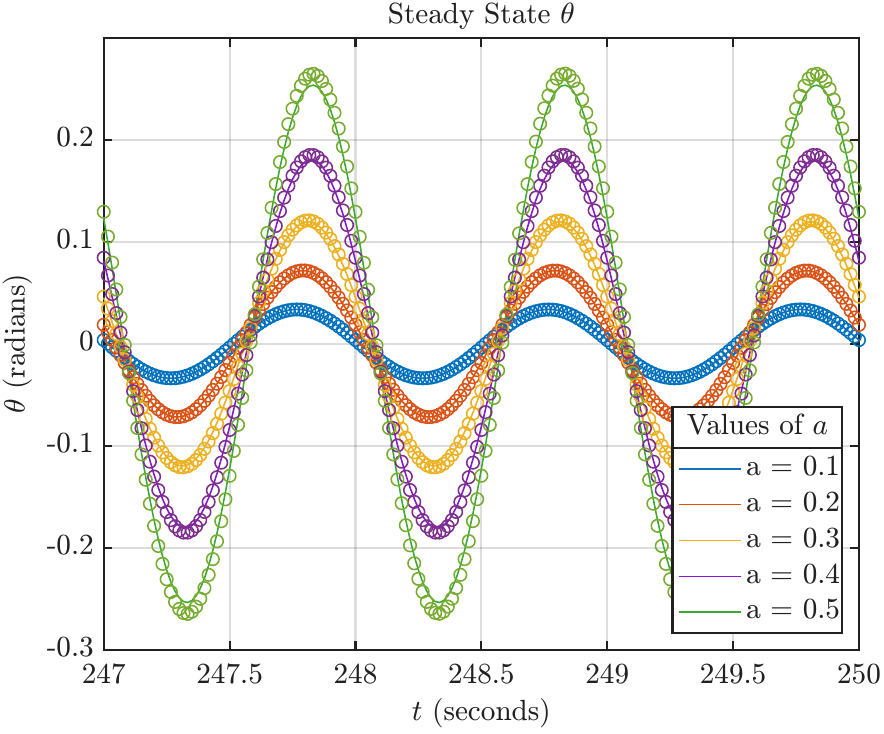}
    \caption{Steady-state orientation angle $\theta$ of the slot-car model over three oscillation periods for parameter group 1 (Table~\ref{tab:parameter_group1}). Circles denote the full ODE simulation; solid lines represent the harmonic balance solution. As input amplitude $a$ increases, both a higher magnitude of $\theta$ and a slight phase lag become apparent. The close agreement between methods confirms that the response is dominated by the first harmonic.}
    \label{fig:varied_a_theta_ss}
\end{figure}

Figure~\ref{fig:varied_a_theta_ss} shows the steady-state orientation angle $\theta$ over three full oscillation periods. Circles represent the full ODE simulation of Eqs. \eqref{eq:accel1}-\eqref{eq:ThetaDot3}, while solid lines indicate the harmonic balance solution of the cubic model. The two methods yield closely aligned results, indicating that the model response is dominated by the fundamental, or first, harmonic, justifying the truncation of higher-order terms in the harmonic balance formulation. Notably, as the input amplitude $a$ increases, the magnitude of $\theta$ grows, and a slight rightward shift in the peak emerges, indicating a slight increase in phase lag. While subtle, this trend could be relevant in contexts where timing and coordination are critical, such as biological control or higher-fidelity modeling.

\begin{figure}[htbp]
    \centering
    \includegraphics[width=\figwidth]{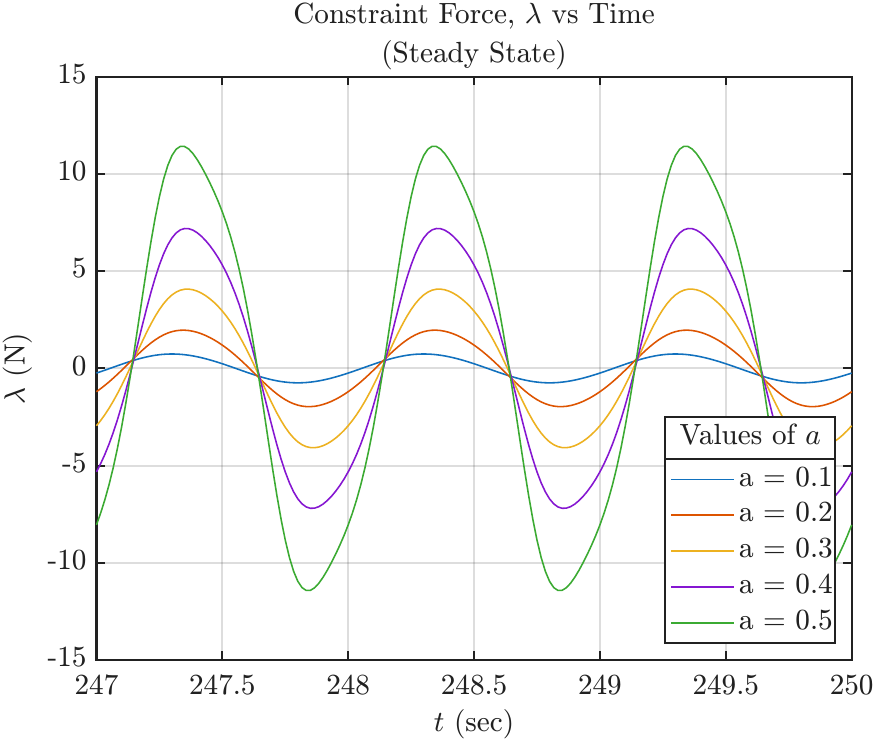}
    \caption{Steady-state constraint force $\lambda$ of the slot-car model over three oscillation periods for parameter group 1 (Table~\ref{tab:parameter_group1}). Each curve corresponds to a different input amplitude $a$. As $a$ increases, the force magnitude grows and waveform distortions emerge, reflecting the nonlinear coupling between body kinematics and the nonholonomic constraint.}
    \label{fig:varied_a_lambda_ss}
\end{figure}

The constraint force $\lambda$, computed from numerical solutions of Eqs. \eqref{eq:accel1}-\eqref{eq:constraintforce1}, is shown in Figure~\ref{fig:varied_a_lambda_ss}. The plot shows three full periods, with curves corresponding to increasing input amplitudes $a$. As $a$ increases, the force amplitude increases, and the waveform begins to distort slightly. 
This distortion reveals nonlinear behavior, which could warrant further investigation in studies focused on large-amplitude motion.

\begin{figure}[htbp]
    \centering
    \includegraphics[width=\figwidth]{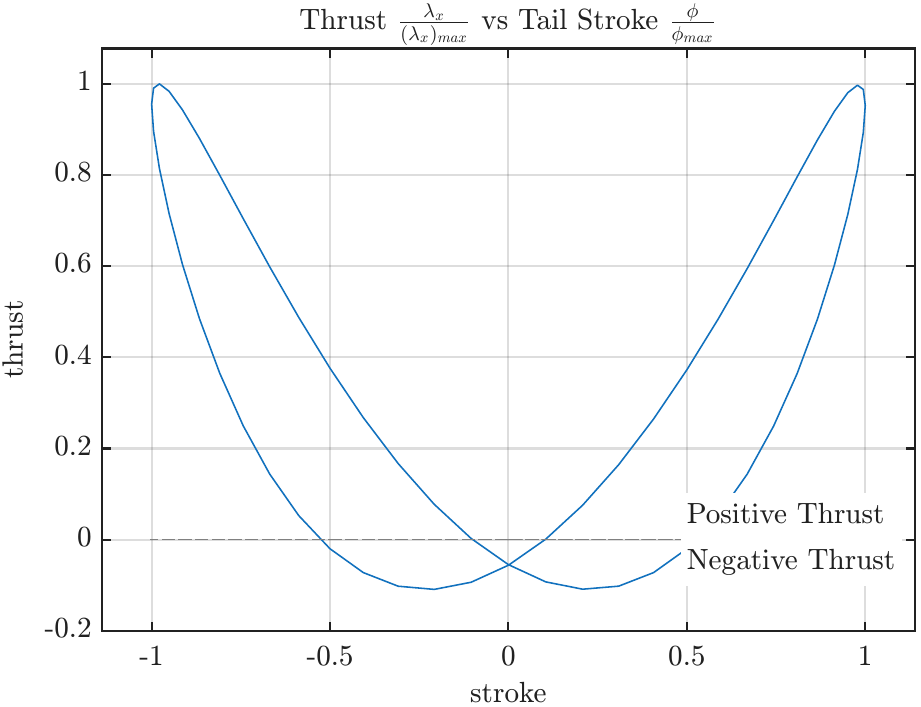}
    \caption{Steady-state thrust in the $x$-direction, $\lambda_x = \lambda \sin(\theta + \phi)$, plotted against the tail angle $\phi$ for parameter group 1 (Table~\ref{tab:parameter_group1}). The resulting figure-eight Lissajous curve reveals a frequency-doubling effect and minimal phase lag, consistent with thrust generation on both sides of the stroke cycle.}
    \label{fig:thrust_vs_stroke}
\end{figure}

The horizontal component of the constraint force, $\lambda_x = \lambda \sin(\theta + \phi)$, represents thrust. Figure~\ref{fig:thrust_vs_stroke} plots $\lambda_x$ against the tail angle $\phi$, revealing a figure-eight Lissajous curve with minimal phase lag.  During the oscillation in the Lissajous figure, the thrust has two oscillations while the tail stroke goes through one oscillation.  That is, the thrust has twice the frequency of the tail stroke.  The frequency doubling occurs due to quadratic nonlinearity in Eqn.~\eqref{eq:accel1}, and physically because the tail generates thrust on each side of the body during a tail stroke—once when sweeping left and once when sweeping right—resulting in two thrust peaks per cycle. The minimal lag between tail motion and thrust production reflects efficient energy transfer. If added fluid mass effects were present, we might expect greater lag and reduced thrust amplitude due to the compliance of the surrounding fluid~\cite{Lighthill1960}.

\subsection{Dependence on Input Amplitude and Frequency}

To further assess the predictive capabilities and limitations of the slot-car model, we examine its behavior under varying excitation conditions. In particular, we investigate how changes in input amplitude $a$ and oscillation angular frequency $\omega$ affect forward velocity, orientation angle, and the constraint force.

\begin{figure}[htbp]
    \centering
    \includegraphics[width=\figwidth]{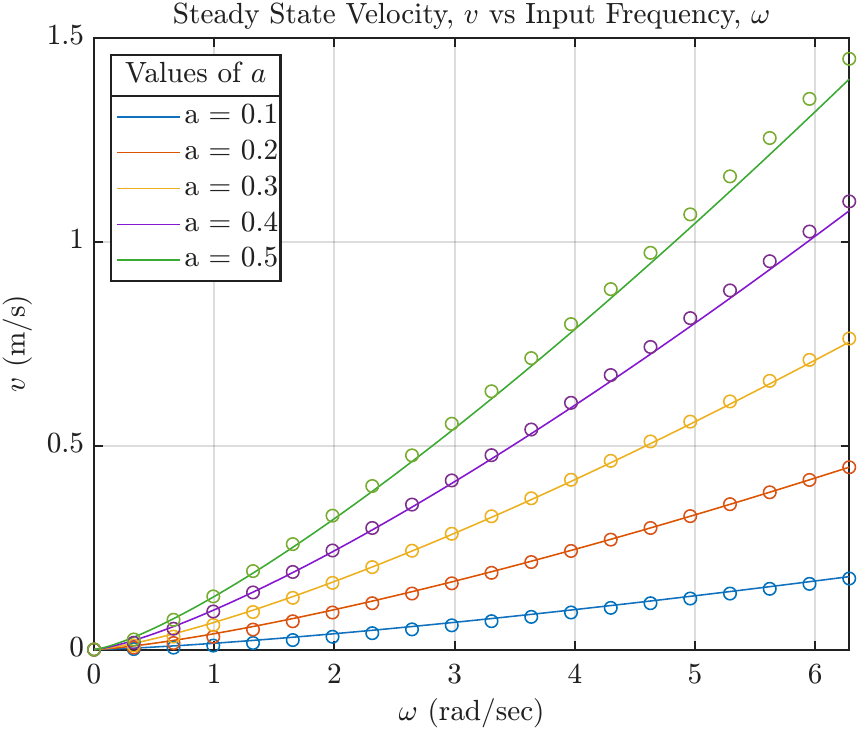}
    \caption{Steady-state mean forward velocity as a function of input angular frequency $\omega$ for five input amplitudes $a$. Solid lines represent harmonic balance solutions; circles indicate full ODE simulation results. Velocity increases approximately linearly with frequency, and higher amplitudes produce disproportionately greater velocities, reflecting mild nonlinear amplification.}
    \label{fig:varied_hbv}
\end{figure}

Figure~\ref{fig:varied_hbv} shows the steady-state mean forward velocity as a function of $\omega$ for five different input amplitudes $a$. Each solid line represents the harmonic balance solution, and the overlaid circles indicate results from full ODE simulations.  The agreement suggests that the cubic model and harmonic balance solution are valid in the parameter range studied. As in Figure~\ref{fig:varied_a_xdot_full}, increasing $a$ leads to higher mean velocity, and across all cases, the velocity increases with frequency. However, the spacing between curves grows with amplitude—indicating that the velocity gain becomes slightly larger at higher amplitudes. This behavior suggests nonlinear amplification of thrust with respect to input amplitude. A similar pattern appears in the steady-state orientation angle $\theta$,  as described next.

\begin{figure}[htbp]
    \centering
    \includegraphics[width=\figwidth]{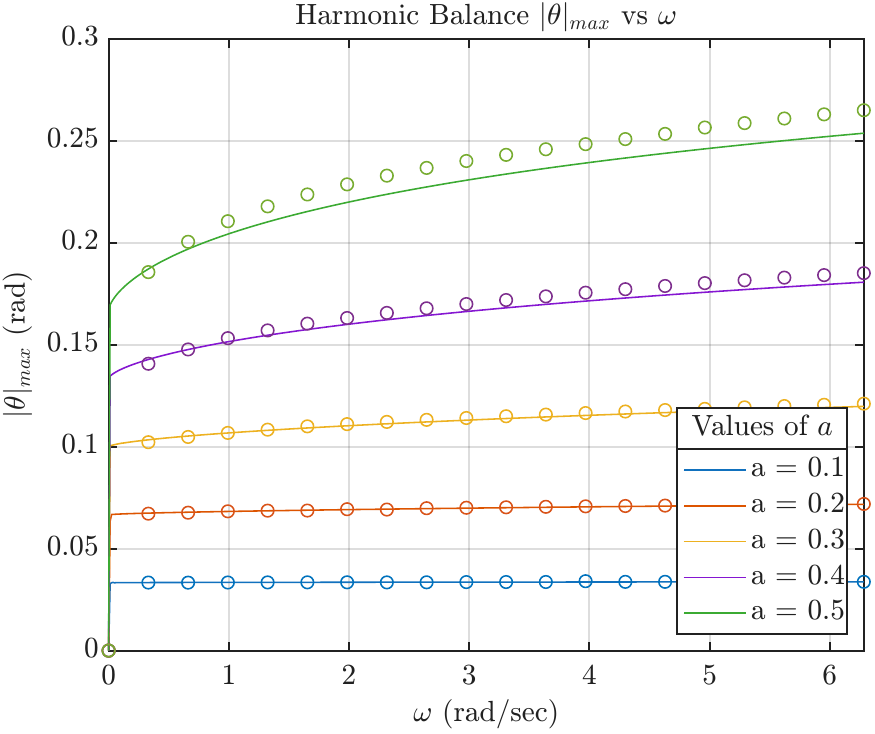}
    \caption{Steady-state orientation amplitude $\theta$ as a function of input angular frequency $\omega$ for five input amplitudes $a$. Solid lines show harmonic balance results; circles indicate full ODE simulations. Frequency dependence is minimal at low amplitudes but becomes more pronounced as $a$ increases, suggesting nonlinear coupling and possible amplification of slot constraint effects.}
    \label{fig:theta_hb_vs_w}
\end{figure}

Figure~\ref{fig:theta_hb_vs_w} shows the steady-state amplitude of the head orientation angle $\theta$ as a function of $\omega$ for five input amplitudes $a$. Solid lines represent harmonic balance results and circles denote full ODE simulations. At low amplitudes, the response remains nearly flat across the frequency range, suggesting that $\theta$ is only weakly influenced by $\omega$ in this regime. However, as amplitude increases, a clear frequency dependence emerges, particularly at $a = 0.5$. This trend indicates nonlinear coupling between tail motion and the constraint dynamics. 

\begin{figure}[htbp]
    \centering
    \includegraphics[width=\figwidth]{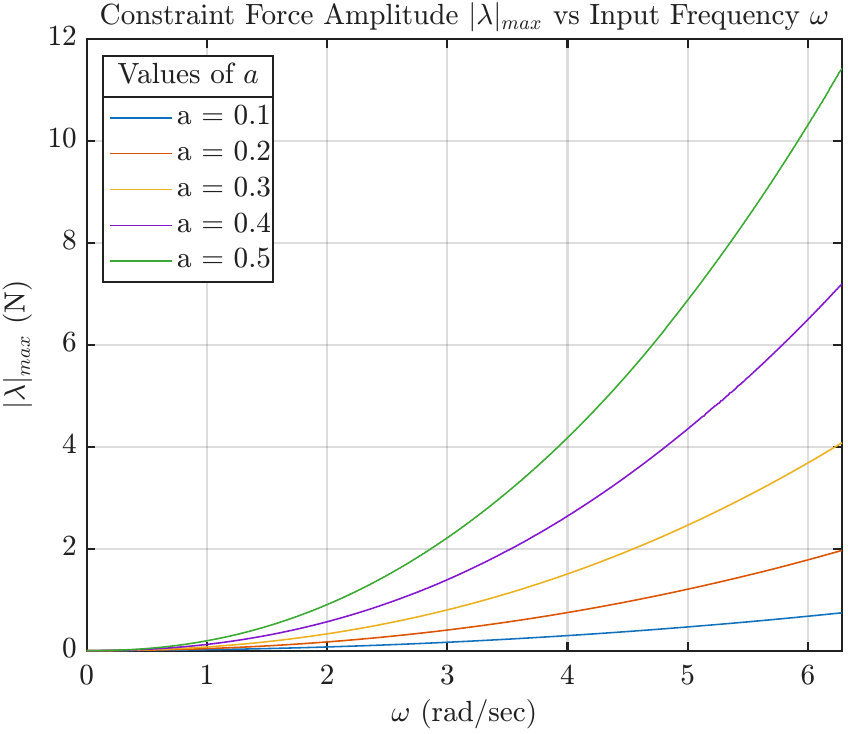}
    \caption{Steady-state maximum constraint force $\lambda_{\max}$ as a function of input frequency $\omega$ for five input amplitudes $a$. The rapid, nonlinear growth at higher amplitudes reflects increasing impulse demand and highlights how the nonholonomic constraint concentrates momentum exchange that, in real swimming, would be distributed across the fluid–body interface.}
    \label{fig:varied_lambda_max}
\end{figure} 

Figure~\ref{fig:varied_lambda_max} shows the steady-state maximum constraint force $\lambda_{\max}$ as a function of $\omega$ for five different input amplitudes $a$. In all cases, $\lambda_{\max}$ increases monotonically with frequency, but this rate of growth becomes nonlinear as amplitude rises. This nonlinear scaling is more pronounced for the constraint force than for the velocity response.  Physically, this suggests an increasing lateral force demand as stroke frequency rises. In the absence of explicit fluid modeling, the nonholonomic constraint must account for all lateral momentum exchange, effectively standing in for the net fluid reaction force.

\section{Model validation based on CFD simulations}
\label{sec:cfd}

In this section, we conduct CFD simulations to validate the assumptions and evaluate the predictive accuracy of our low-order model developed in Sec.~\ref{sec:model_eq}. The primary objectives of this validation study are to address several key questions: (i) Does a nonholonomic constraint actually exist in the swimming dynamics of a two-body system? (ii) If it does exist, where is it? (iii) How do variations in fin size and location on Body 2 affect the constraint characteristics and overall swimming dynamics?

First, we establish our CFD methodology, including governing equations, simulation setup, and parameter ranges (Sec.~\ref{sec:simulation-methodologies}). We then justify key assumptions of our low-order model by analyzing the existence and location of the nonholonomic constraint, as well as the center of pressure behavior from the CFD results (Sec.~\ref{sec:justification-simulation-setup}). Finally, we validate the model's predictive accuracy by comparing with CFD data for velocity, body orientation, and constraint force, while examining the effects of fin geometry variations (Sections~\ref{sec:validation}). 

In the following analysis, we adopt standard CFD notation where $\vec{u}$ (or $u_i$ in index notation and individual components $u$ and $v$) denotes the swimming velocity, $\Omega$ represents angular velocity, and the total force is denoted as $\vec{F}$. The normal force $F_n$ on the fin corresponds to the magnitude of the constraint force $\lambda$ from the dynamics model.

\subsection{Simulation methodologies}
\label{sec:simulation-methodologies}

\subsubsection{Governing equations}

 \begin{figure*}  
 \begin{center}
 \includegraphics[width=1\linewidth]{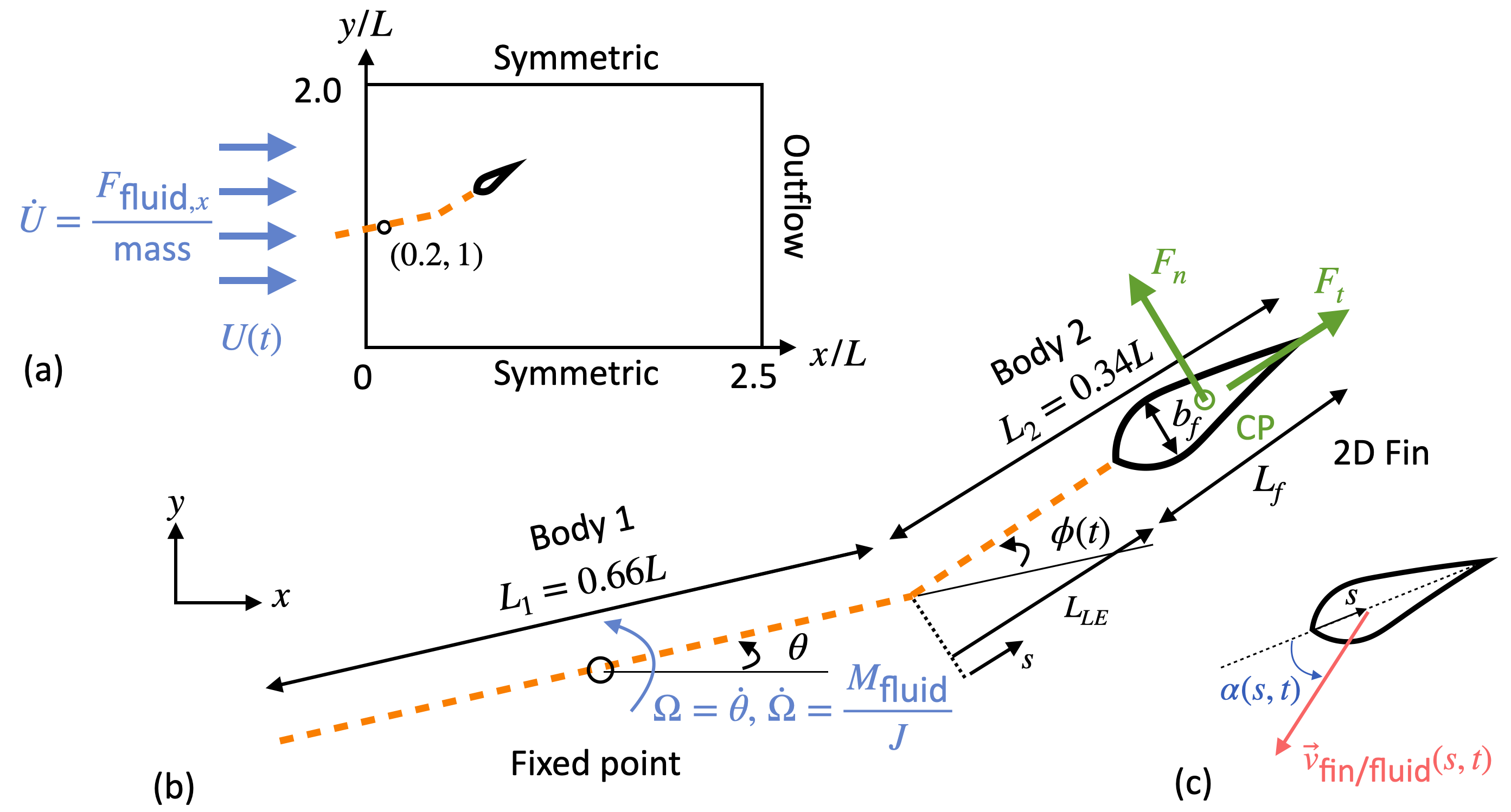}
  \caption{Setup of FSI simulation of self-propelled slot-car fish swim of the two-body system. Boundary conditions and domain size are shown in (a). A sketch of slot-car is shown in (b). Body 1 is anchored at the fixed point. Tail (Body 2) ``flapping'' motion ($\phi(t)$) is imposed. The inflow velocity $U(t)$ and body rotation angle $\theta(t)$ are obtained from simulations, depending on the total force $F_\text{fluid,x}$ and moment around the fix point $M_\text{fluid}$. $L$ is the total length of the two-link body; $L_1$ and $L_2$ are respective lengths of the Bodies 1 and 2. (c)  Definition of the angle of attack, $\alpha$, based on fin orientation and the velocity of fin related to far-field fluid, $\vec{v}_\text{fin/fluid}$.}
  \label{fig:sketch}	
 \end{center}
\end{figure*}

Two-dimensional (2D) simulations are carried out for a two-body system equivalent to that examined in Sec.~\ref{sec:model_eq} under self-propelled swim in a stationary flow. A sketch is shown in Figure~\ref{fig:sketch}. The flapping motion of Body  2 (i.e. the tail) was imposed by prescribing the relative angle $\phi(t)$ as a harmonic oscillation. Both  linear and angular accelerations of the body system as a result of the flapping are simulated.  

A few simplifications are made to lower simulation cost and solver complexity.
First, the exact solution of a two-body system under arbitrary motion in fluid would require a non-inertial frame of reference attached to the body system \citep{Borazjani08thesis,KimKC06}, which needs non-trivial modification of a fluid solver that was built originally for an inertial frame of reference. Here,  the slot car is assumed to undergo steady swim at all times, with the far-field velocity at the domain inlet allowed to vary in time depending on the total fluid force component $F_x$  \citep[following the approach of][]{HanchiHBMOKL13}. This simplification allows the use of the inertial frame of reference attached on the body system. The solution is correct at the steady-swim state, while incorrect during the transient state; only the steady state results are analyzed herein.
In addition, since for fish most of the fluid force is generated on the fin, only the fin of the body-fin system is resolved in the solid representation. For Body 1 and Body 2 (except for the fin), the effect of the body on the fluid flow is ignored but the mass and the moment of inertia of both bodies were considered in the fluid-solid coupling. This simplification is consistent with the  low-order model assumptions that fluid force is applied at the fin only.
This assumption corresponds to the flow around a large aspect ratio quasi-two-dimensional fin (such as that of a tuna).  The fin is simulated as a rigid hydrofoil   \citep[of the form used by][]{HanchiHBMOKL13}. Other fin geometry may also be used; the main requirement is a tapered trailing edge to reduce the form drag, to approximate fish swim.

As the coordinate system is fixed on the body system, the ``fish'' is anchored (in $x$ and $y$) at the midpoint of Body 1, consistent with the low-order slot-car model where the lateral ($y$) motion of the anchored point is restricted. Here, $x$ and $y$ are the  fluid flow direction (relative to the body system) and the transverse direction.
The inlet velocity $U$ of the fluid flow is adjusted to the total force from the body to the fluid, $F_x$, according to Newton's second law applied to the two-body system,
\begin{equation}
	\label{e:dyn1}
	\frac{dU}{dt} = -\frac{F_x}{m},
\end{equation}
where $m$ is the slot-car mass. The angular motion of the body system is determined by 
\begin{equation}
	\label{e:dyn2}
	\frac{d\Omega}{dt} = \frac{M}{J},
\end{equation}
where $\Omega$, $M$ and $J$ are the angular velocity, total fluid force moment, and slot-car moment of inertia with respect to the anchored point. 
 
The simulations use a finite-difference code that solves the unsteady  incompressible  Navier-Stokes equations. The fluid boundary conditions on the moving solid body are imposed using an immersed boundary method based on the volume-of-fluid approach \citep{YuanP14b,YuanP14}.
The governing equations in the index notation are:
\begin{eqnarray}
  \label{eq:continuity_filtered_dns}
  \dpr{{u}_i}{x_i} & = & 0, \\
  \label{eq:momentum_filtered_dns}
  \dpr{{u}_j}{t} 
  +\dpr{{u_i}{u}_j}{x_i} & = & 
  -\dpr{{P}}{x_j}+\nu\nabla^2{u}_j 
  +  f_j,
\end{eqnarray}
where 
$P=p/\rho$ is the modified pressure, $\rho$ the density and $\nu$ the kinematic viscosity. The term $f_j$ in Equation~(\ref{eq:momentum_filtered_dns}) is a body force required by the immersed boundary method to impose non-slip/penetration boundary conditions at the fin.
Equations~(\ref{eq:continuity_filtered_dns}) to (\ref{eq:momentum_filtered_dns}) are solved on a staggered grid using second-order central differences for all terms and the second-order accurate Adams-Bashforth semi-implicit time advancement.
The solid governing equations~(\ref{e:dyn1}) and (\ref{e:dyn2}) are  discretized based on the Euler explicit time advancement.  The fluid-structure coupling uses a weak coupling approach where fluid forces are computed from the converged flow field at each time step, then applied to update solid motion for the next time step.

\subsubsection{Simulation setup}

\begin{table*}
    \centering
    \begin{tabular}{lccccc}
        \hline
        \textbf{Case} & \textbf{Re$_L$} & \textbf{St} & \textbf{$T$} & \textbf{$a$} & \textbf{Resolution} \\
        \hline
        \textbf{Case 1:} $L_f/L_2 = 1$, $L_{LE}/L_2 = 0$ & \textbf{7.05E+3} & \textbf{0.73} & \textbf{1.0} & \textbf{0.20} & \textbf{R2} \\
        \hline
        \multicolumn{6}{l}{\textbf{Case 2:} $L_f/L_2 = 0.5$, $L_{LE}/L_2 = 0.12$} \\
        \multicolumn{6}{l}{Varying resolution:} \\
        case2\_t10\_a02\_re3 & 5.70E+3 & 0.90 & 1.0 & 0.20 & R1 \\
        \textbf{case2\_t10\_a02\_re3} & \textbf{6.26E+3} & \textbf{0.82} & \textbf{1.0} & \textbf{0.20} & \textbf{R2} \\
        case2\_t10\_a02\_re3 & 6.39E+3 & 0.80 & 1.0 & 0.20 & R3 \\
        \hline
        \textbf{Case 3:} $L_f/L_2 = 0.5$, $L_{LE}/L_2 = 0.25$ & \textbf{7.73E+3} & \textbf{0.67} & \textbf{1.0} & \textbf{0.20} & \textbf{R2} \\
        \hline
        \multicolumn{6}{l}{\textbf{Case 4:} $L_f/L_2 = 0.5$, $L_{LE}/L_2 = 0.5$} \\
        \multicolumn{6}{l}{Varying Reynolds number:} \\
        case4\_t10\_a02\_re1 & 6.59E+2 & 1.95 & 1.0 & 0.20 & R2 \\
        case4\_t10\_a02\_re2 & 2.80E+3 & 0.92 & 1.0 & 0.20 & R2 \\
        \textbf{case4\_t10\_a02\_re3} & \textbf{9.28E+3} & \textbf{0.58} & \textbf{1.0} & \textbf{0.20} & \textbf{R2} \\
        \hline
        \multicolumn{6}{l}{Varying amplitude:} \\
        case4\_t10\_a015\_re3 & 5.33E+3 & 0.96 & 1.0 & 0.15 & R2 \\
        \textbf{case4\_t10\_a02\_re3} & \textbf{9.28E+3} & \textbf{0.58} & \textbf{1.0} & \textbf{0.20} & \textbf{R2} \\
        case4\_t10\_a03\_re3 & 1.65E+4 & 0.31 & 1.0 & 0.30 & R2 \\
        case4\_t10\_a04\_re3 & 2.22E+4 & 0.23 & 1.0 & 0.40 & R2 \\
        \hline
        \multicolumn{6}{l}{Varying period:} \\
        case4\_t05\_a02\_re3 & 2.30E+4 & 0.45 & 0.5 & 0.20 & R2 \\
        \textbf{case4\_t10\_a02\_re3} & \textbf{9.28E+3} & \textbf{0.58} & \textbf{1.0} & \textbf{0.20} & \textbf{R2} \\
        case4\_t20\_a02\_re3 & 2.83E+3 & 0.91 & 2.0 & 0.20 & R2 \\
        case4\_t40\_a02\_re3 & 6.76E+2 & 1.90 & 4.0 & 0.20 & R2 \\
        \hline
    \end{tabular}
\caption{Summary of simulation cases with varying parameters. The case naming convention follows the format ``caseX\_tYY\_aZZ\_reW'', where X denotes the case number (1-4), YY indicates the dimensionless tail-beat period ($T$, used as the reference time scale), ZZ represents the tail-beat amplitude ($a$ in radians), and W indicates the Reynolds number group (1-3). Three spatial resolutions are considered: R1 ($1000 \times 700$), R2 ($1600 \times 1100$), and R3 ($2000 \times 1500$), with minimum grid sizes $\Delta_{min}/L$ of 0.0012, 0.0007, and 0.0003, respectively. $L_f/L_2$ indicates the fin length relative to Body 2 length, and $L_{LE}/L_2$ represents the location of the fin leading edge measured from the start of Body 2. Re$_L$ is the Reynolds number based on $L$ and steady-swim speed $U_s$. St is the Strouhal number, defined as St$ = f(2A_m)/U_s$, where $A_m$ is the maximum lateral excursion of the tail. Rows highlighted in bold are the baseline cases, to be compared in Fig.~\ref{fig:aoa_dist_cases} to compare the fin configurations.}    \label{tab:simulation_cases}
\end{table*}

Four cases with different dimensionless fin lengths ($L_f/L_2$) and dimensionless locations of the fin leading edge ($L_{LE}/L_2$, measured from the head of Body 2) are tested; they are listed in Table~\ref{tab:simulation_cases}. 
Case 1 uses a fin that is as long as Body 2, while Cases 2 to 4 have the same fin length of $L_2/2$ but are different in fin locations, with the fin leading edge progressively moving down Body 2, from 12\% to 50\% locations, measured from the start of Body 2. 
Case 4 is shown in Sec.~\ref{sec:justification-simulation-setup} to be the optimal configuration that produces a time-averaged location of the minimum angle-of-attack that is closest to the imposed NH location in the low-order model (i.e. midpoint of Body 2), as well as a relatively close match of the center of pressure (CP) location.
In Table~\ref{tab:simulation_cases}, sub-cases with different simulation parameters (tail-beat period, maximum tail-beat angle, fluid viscosity, and resolution) under each case category are also listed. The resultant Reynolds number and Strouhal number measured at the steady-swim state are tabulated. Note that the fluid flow (around the fin) is laminar in all cases due to the low Reynolds number based on the fin length.

The dimensionless  $m$ and $J$ (based on fluid density and slot-car length) match those used in the low-order model: $m/(\rho_f L^3)=0.068$ and $J/(\rho_f L^4) = 0.003$.  
An arbitrary initial inlet velocity value of $1L/T$ is imposed; other values would give the same steady-swim results. The steady-swim velocity $U_s$ is measured from the simulation results after the steady swim state is reached. The Reynolds number and Strouhal number are calculated as
\begin{equation}
	\mathrm{Re}_{L}= U_s L/\nu,
\end{equation}
\begin{equation}
	\mathrm{St} = f (2A_m)/U_s,
\end{equation}
where $A_m\equiv L_2\max[\sin(\phi)]$ is the width of the wake quantified by the maximum lateral excursion of the tail over a cycle \citep{BorazjaniBS09}. The Reynolds number represents the ratio of inertial to viscous forces, while the Strouhal number represents the ratio between the velocity of fin oscillation and that of the swim velocity.

\begin{figure*}  
 \begin{center}
 \includegraphics[width=.6\linewidth]{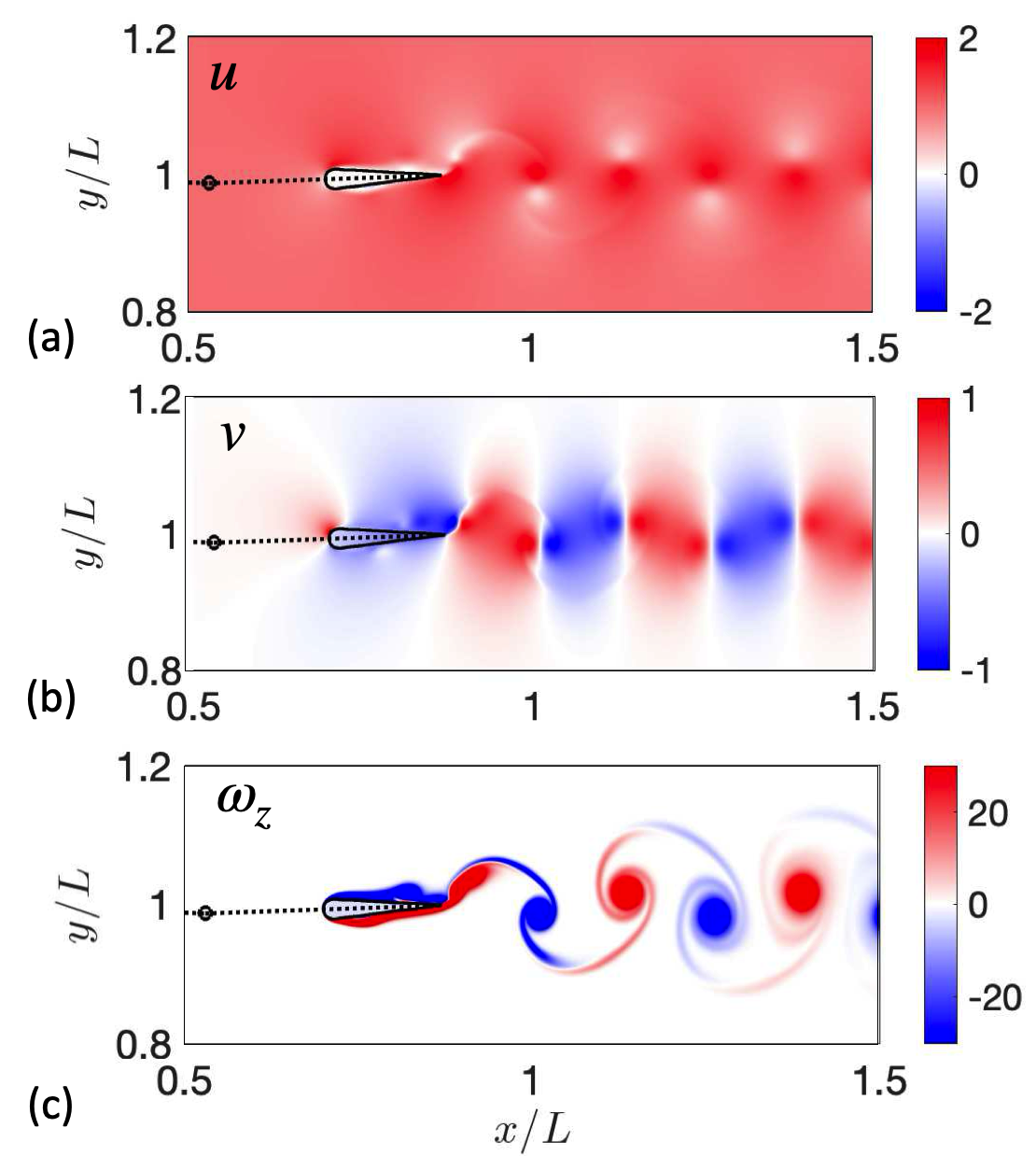}
  \caption{CFD results shown by instantaneous contours of fluid properties in a near-fin region, at steady-swim state for \textit{case4\_t10\_a02\_re3}: (a) streamwise fluid velocity, (b) transverse fluid velocity, and (c) vorticity. All quantities are normalized by $T$ and $L$. $\dotted$ Bodies 1 and 2 (Body 1 extends out of the field of view), connected at location marked by~$\circ$.}
  \label{fig:contours_slotcar}	
 \end{center}
\end{figure*}

Reynolds number effects on FSI were tested for Case 4 (see \ref{sec:A-re}), showing that body kinematics become Reynolds-number-independent above a threshold value. All simulations are conducted at the highest Reynolds number tested (``re3'' in Table~\ref{tab:simulation_cases}) to eliminate Reynolds number effects, though the flow remains laminar as shown by the vorticity field in Fig.~\ref{fig:contours_slotcar}(c). A grid refinement study for Case 2 confirmed that the mesh with $(1600,1100)$ grid points (minimum grid size $\Delta_{\min}/L = 0.0007$) provides grid-independent results.
Finally, various tail-beat periods $T=1/f$ (dimensionless value from 0.5 to 4) and amplitudes $a$ (from 0.1 to 0.4 in radians) are used for Case 4 to validate the model.

 \begin{figure*}  
 \begin{center}
 \includegraphics[width=\linewidth]{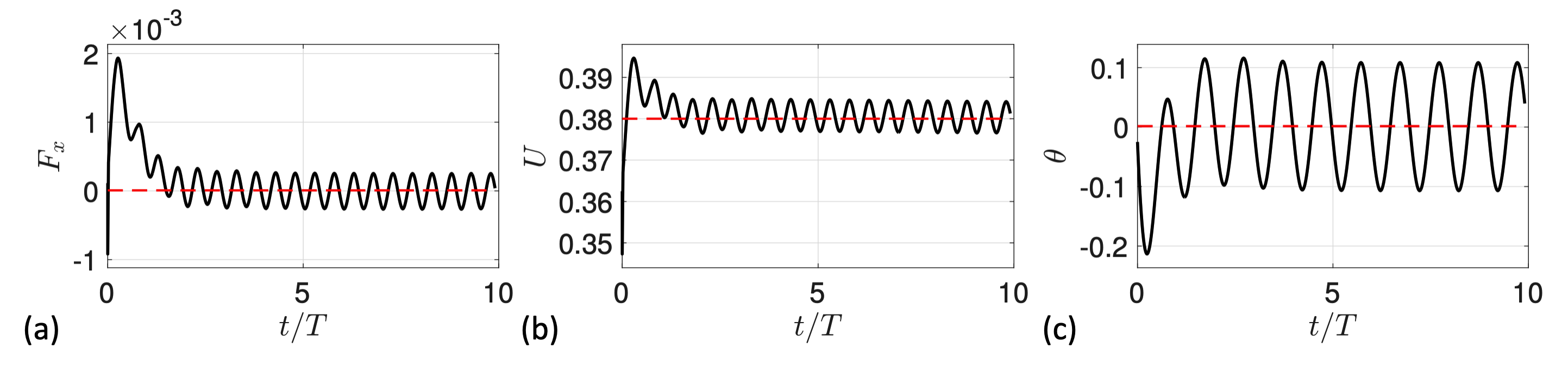}
  \caption{Transient development of various quantities in a CFD simulation: (a) total streamwise fluid force $F_x(t)$ on the fin, (b) swim velocity $U(t)$, (c) Body-1 orientation angle, $\theta$.  \textit{case4\_t10\_a03\_re3}  is used, with initial condition from the steady state of another case.  {\color{red}$\dashed$} Long-time average values. }
  \label{fig:transient}	
 \end{center}
\end{figure*}
 
Figure~\ref{fig:transient} shows  time variations in Case \textit{case4\_t10\_a02\_re3} of various quantities characterizing the FSI, including the  total fluid force $F_x$ (positive for thrust), instantaneous swim velocity $U(t)$, and orientation angle $\theta$ of Body 1. The initial condition is taken from another case with a steady-swim velocity of around 0.35. 
At early times the tail flapping leads to thrust ($F_x>0$), increasing the swim velocity toward the long-term average of around 0.38. In this period, the body motion is asymmetric (with $\theta$ biased toward negative values). 
At large times ($t/T \approx 2$), the steady swim state is reached. $F_x$ fluctuates around its long-term average of zero, indicating steady swim. $\theta$ also fluctuations around the long-time average value of zero, showing a symmetric response.

Figure~\ref{fig:contours_slotcar} shows instantaneous contours of velocities (streamwise $u$ and transverse $v$ components) and the vorticity ($\omega_z$) at the steady-swim state for  \textit{case4\_t10\_a02\_re3}. 
The $\omega_z$ contour in Fig.~\ref{fig:contours_slotcar}(c) shows that eddies shed from the trailing edge of the fin, generating a high-speed jet downstream of the fin (see Fig.~\ref{fig:contours_slotcar}(a)). This phenomenon creates a thrust that balances the fluid drag force.  
The boundary layer on both sides of the fin is unsteady and non-equilibrium, characterized by local separation and reattachment (Fig.~\ref{fig:contours_slotcar}(c)).

\subsection{Justification of simulation setup}
\label{sec:justification-simulation-setup}

The fin size and fin location are varied across four cases to identify configurations that best resemble the theoretical model scenario. This section provides quantitative justification that the simulated problem is a reasonable representation of the assumptions made in the low-order model by comparing flow characteristics and essential swimming parameters across these cases. One of the cases (Case 4) that demonstrate the best agreement with model parameters is selected for detailed comparison in Sec.~\ref{sec:validation}. The evaluated model assumptions include (i) existence of an NH constraint and its placement at the mid-body-2 location, as well as (ii) a location of fluid-force center of pressure (CP) at the mid-body-2 location, used to calculate the normal force on the fin  (i.e. the constraint force $\lambda$) from the FSI solution of the low-order model in Sec.~\ref{sec:model_eq}.

\begin{figure*}
\centering
\includegraphics[width=\linewidth]{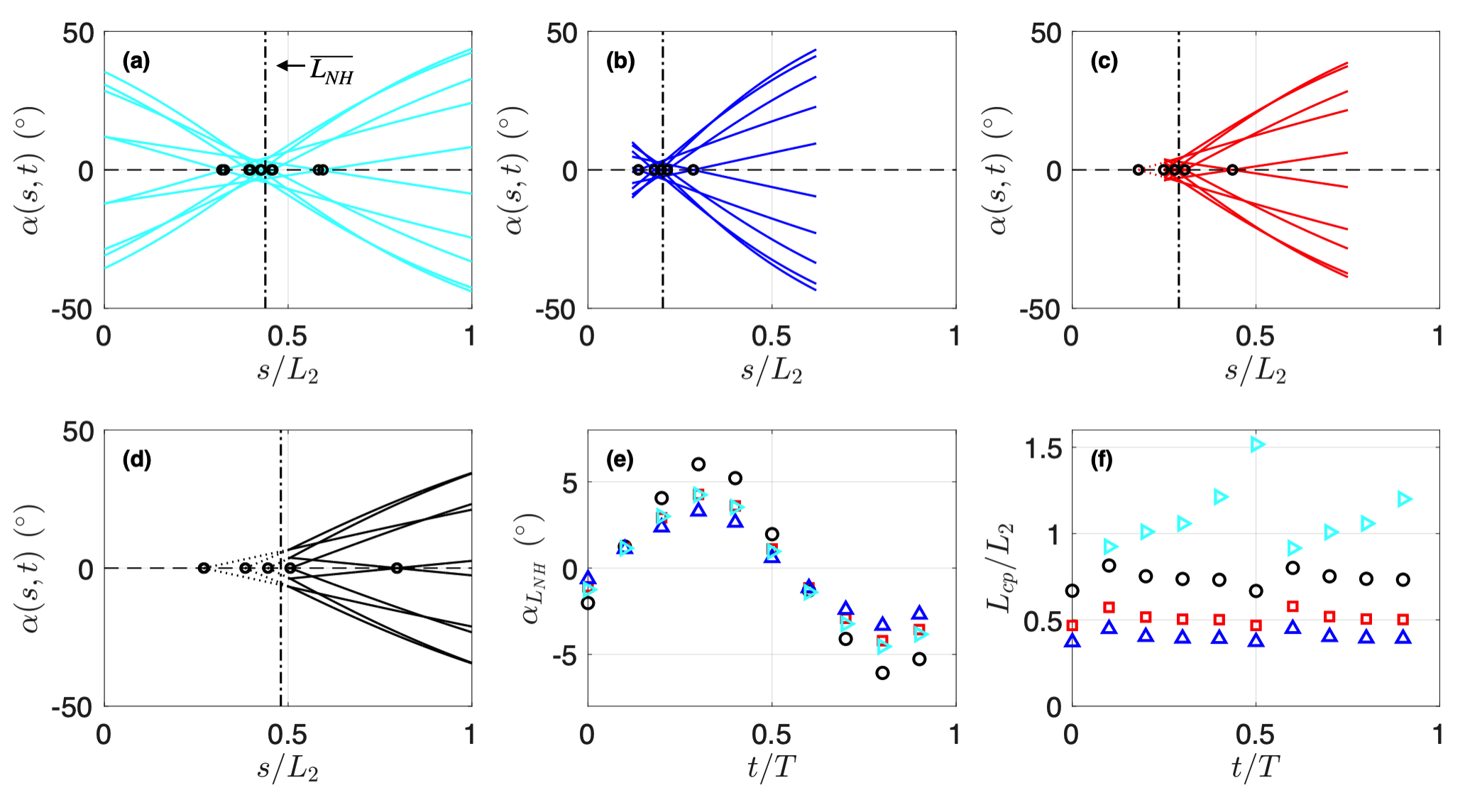}
\caption{  Effects of fin length and location used in simulations, comparing Cases 1 to 4. (a-d) Instantaneous distributions of $\alpha$ on fin along $s$ on Body 2, at 10 equally spaced time instances during one tail-beat cycle. Cases 1 to 4 are  in (a) to (d), respectively. $\circ$ Instantaneous zero-AOA locations ($L_{NH}$); $\chndot$ (vertical): the period-averaged NH location ($\overline{L_{NH}}$). (e) Instantaneous AOA at averaged NH location. (f) Instantaneous center of pressure (CP) location. In (e,f), cases are colored same as in (a-d).   }
\label{fig:aoa_dist_cases}
\end{figure*}

To evaluate the applicability using an NH constraint to model swimming, the instantaneous distribution of the angle-of-attack (AOA, denoted as $\alpha$) of the fin is shown in Fig.~\ref{fig:aoa_dist_cases}(a-d) as a function of location $s$ ($s=0$ at fin leading edge) along the full length of Body 2, at  ten equally spaced time instances within one period, for Cases 1-4, respectively. AOA is defined as the angle between the local fin chord and the fin velocity relative to that of the far-field fluid flow (see Fig. \ref{fig:sketch}(c)). As the local fin velocity varies with $s$, $\alpha$ is a function of $s$.
It is also shown in all cases that $\alpha$ varies significantly in time. However, all cases display an $s$ location near which $\alpha$ is close to zero at all times, indicating that a NH constraint approximately applies.

To quantitatively evaluate the applicability of the exact NH constraint, we define the location of the exact instantaneous NH location as the instantaneous $s$ location at which $\alpha(s,t)=0$,  quantified by the length $L_{NH}(t)$ (i.e. the distance from the start of Body 2 to the instantaneous zero-AOA location) and marked by black circles in Fig.~\ref{fig:aoa_dist_cases}(a-d). We also define an equivalent NH location, as the period-averaged location of the zero-AOA point, denoted by the length $\overline{L_{NH}}$ and marked by the vertical dashed line in Fig.~\ref{fig:aoa_dist_cases}(a-d).
Results show that, for all cases, there is significant variation of $L_{NH}(t)$, indicating that a fixed, exact NH location does not exist. In addition, Cases 1 and 4 yield a time-averaged zero-AOA location ($\overline{L_{NH}}$) closest to the mid-point of Body 2, as assumed in the low-order model.
Note that, at some time instances $\alpha(s)$ does not reach zero throughout the entirety of the fin (i.e. Body 2); at these time instances, the extrapolated zero-crossing location of $\alpha(s)$ is used as the instantaneous NH location.
Figure~\ref{fig:aoa_dist_cases}(e) shows that the instantaneous $\alpha(t)$  at the averaged NH location is small, less than 5 degrees in magnitude for all cases. 
These observations indicate that the setups of Cases 1 and 4  are overall consistent with the model assumption of the existence of an NH constraint at mid Body 2.

Figure~\ref{fig:aoa_dist_cases}(f) shows the instantaneous center of pressure (CP) locations for all cases throughout a period. The CP location is measured by $L_{cp}$ -- the length from the start of Body 2 to the CP location.  The CP location is calculated based on the moment of forces around the leading edge of Body~2,
\begin{equation}
\label{e:Lcp}
L_{cp}=\frac{\int_{o}^{L_{2}} s f_{n} ds }{\int_{o}^{L_{2}} f_{n} ds},
\end{equation}
where $f_n$ is the distributed fluid force per unit fin length. 
Cases 2 and 3 give CP locations that overall approximate the assumption of a mid-point CP location. Case 4 gives a value that is predominantly at $0.75 L_2$.
In Case 1, CP is located near the end of Body 2 and often reaches locations outside Body 2 ($L_{cp}/L_2>1$). This is because of the movement of the region near the head of Body 2 that is toward the opposite direction from that of the rest of the Body 2. The resulting local $f_n$ of a different sign contributed to a lower magnitude of the resultant normal force $F_n=\int_{o}^{L_{2}} f_{n} ds$ and the total moment $\int_{o}^{L_{2}} s f_{n} ds$, but for a smaller proportion for the latter, due to the small $s$ values at locations near the head of this body.

Analysis reveals that, among the four fin setups currently tested, no single case exactly matches all assumptions of the idealized low-order model. This is not surprising, as the model is inherently idealized and is not expected to exactly match the actual swimming motion. However, matching the NH location is more critical than matching the CP location, as this constraint is fundamental to the model formulation and directly determines the body deformation and steady-swim velocity resulting from fluid-structure interaction. While a mismatch in CP location affects the calculation of the constraint force at the fin, it does not influence the model-predicted motion. In the following analysis, we compare all cases with the model predictions while focusing on CFD results of Case 4, as it provides a very good match of equivalent NH location and a CP location not too far from the mid Body 2 location. The mismatch in the CP location remains one possible source of fin-normal force discrepancy between Case 4 simulation and the model prediction.

\subsection{Validation of model results}
\label{sec:validation}

\begin{figure}[htbp]
\begin{center}
\includegraphics[width=1\linewidth]{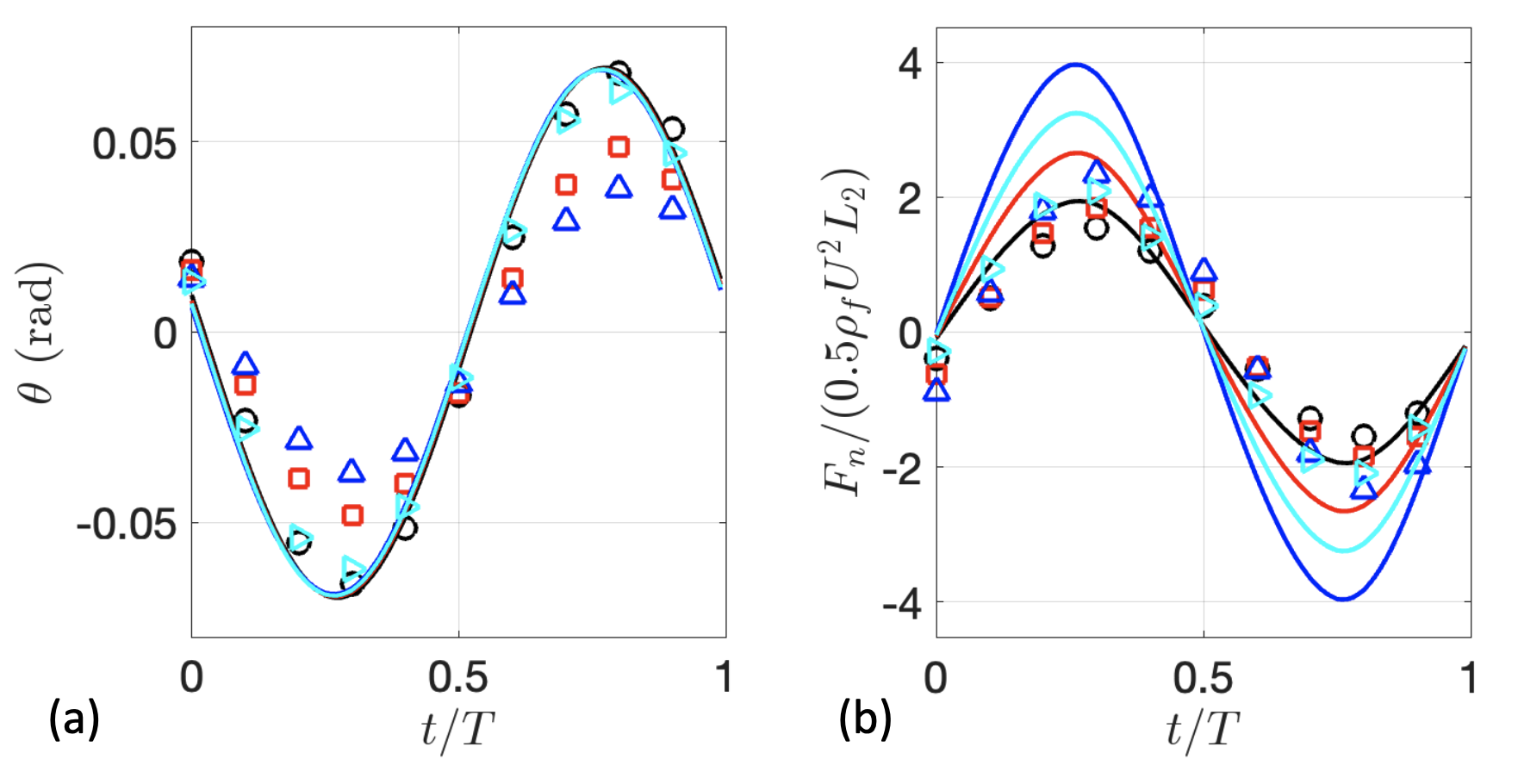}
\caption{Comparison of CFD simulations and model predictions for all four baseline cases in Table~\ref{tab:simulation_cases}: (a) Body 1 orientation angle $\theta$ and (b) dimensionless normal force $F_n$ on the fin. Symbols are CFD results: {\color{cyan}$\triangleright$} Case 1, {\color{blue}$\triangle$} Case 2, {\color{red}$\square$} Case 3, $\circ$ Case 4. Lines are model predictions with drag force calibrated using respective steady-swim velocities from each case, plotted using the same color as the symbols for each case.
}
\label{fig:cfd_results_4cases}
\end{center}
\end{figure}

The comparison of $\theta$ and $F_n$ between CFD simulations and model predictions across all four cases is presented in Fig.~\ref{fig:cfd_results_4cases}, at the baseline values of $T=1.0$, $a=0.20$, and the fluid viscosity associated with the Reynolds number group ``re3''.
Here, the only difference between the model runs is the drag force coefficient (i.e. $a$ in the $C_D$ expression), which is calibrated using the actual steady-swim velocity measured from CFD for each case. The calibration error leads to less than 2\% of the steady-state swim velocity value for all cases.
Figure~\ref{fig:cfd_results_4cases}(a) shows that Cases 1 and 4 exhibit the best agreement between CFD and model prediction in the orientation angle $\theta$; this is consistent with their NH constraint locations closely matching the model assumption of mid-Body 2 placement. 
In addition, Fig.~\ref{fig:cfd_results_4cases}(b) shows that  Case 4 also exhibits an excellent match with CFD in the normal force $F_n$ on the fin, compared to the significant discrepancy in Case 1. This confirms that correct CP location specification in the model is important for accurately predicting the constraint force.
Cases 2 and 3 show poorer agreement with the model predictions in $\theta$ because their NH locations do not align with the mid-Body 2 assumption. Since the NH location fundamentally affects the constraint dynamics and subsequent force generation, these cases cannot be expected to match the force accurately either.
These observations provide evidence that, when the low-order model is configured with geometric parameters that match the CFD simulation, the model can give very good predictions in both body deformation and fluid forces.

\begin{figure}[htbp]  
\begin{center}
\includegraphics[width=1\linewidth]{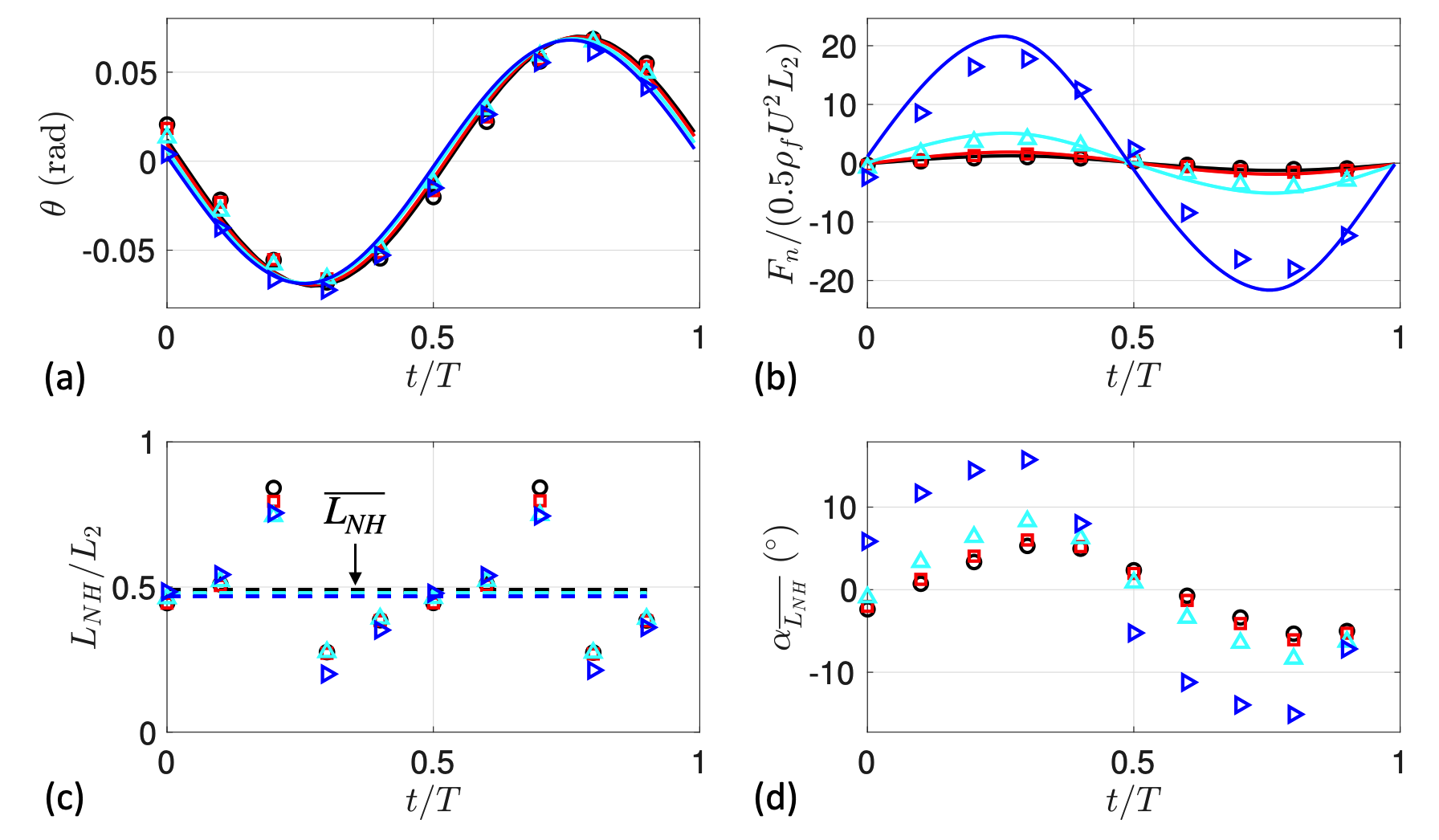}
\caption{Comparison between Case 4 simulations (symbols) and model predictions ($\solid$), with varying flapping period $T$: (a) $\theta$ and (b) $F_n$.
Comparison of (c) NH location and (d) angle of attack at the average NH location among simulation cases.
 $T^*=0.5$ ($\circ$), 1 ({\color{red}$\square$}) 2 ({\color{cyan}$\triangle$}) and 4 ({\color{blue}$\triangleright$}).
}
\label{fig:varT}	
\end{center}
\end{figure}

Next, the comparison between model and CFD results is carried out for a range of values of two swim parameters: tail-beat period $T$ and tail-beat amplitude $a$, focusing on Case~4. 
Figure~\ref{fig:varT} compares simulation and model results for various flapping periods $T$ (i.e., $1/f$). Figure~\ref{fig:varT}(a) shows that the Body 1 deformation is not sensitive to a change in frequency $f$. Similarly, the frequency change does not affect significantly the average NH location  (Fig.~\ref{fig:varT}(c)). The maximum magnitude of the instantaneous AOA  at the averaged NH location varies between  10 to 20 degrees (Fig.~\ref{fig:varT}(d)), increasing with a longer period. This indicates that a higher flapping frequency yields a flow that is slightly closer to the NH constraint. For all values of flapping period, model predictions of $\theta$ and $F_n$ are very close to the CFD results (Fig.~\ref{fig:varT}(a,b)), both in value and in phase.

\begin{figure}  
\begin{center}
\includegraphics[width=1\linewidth]{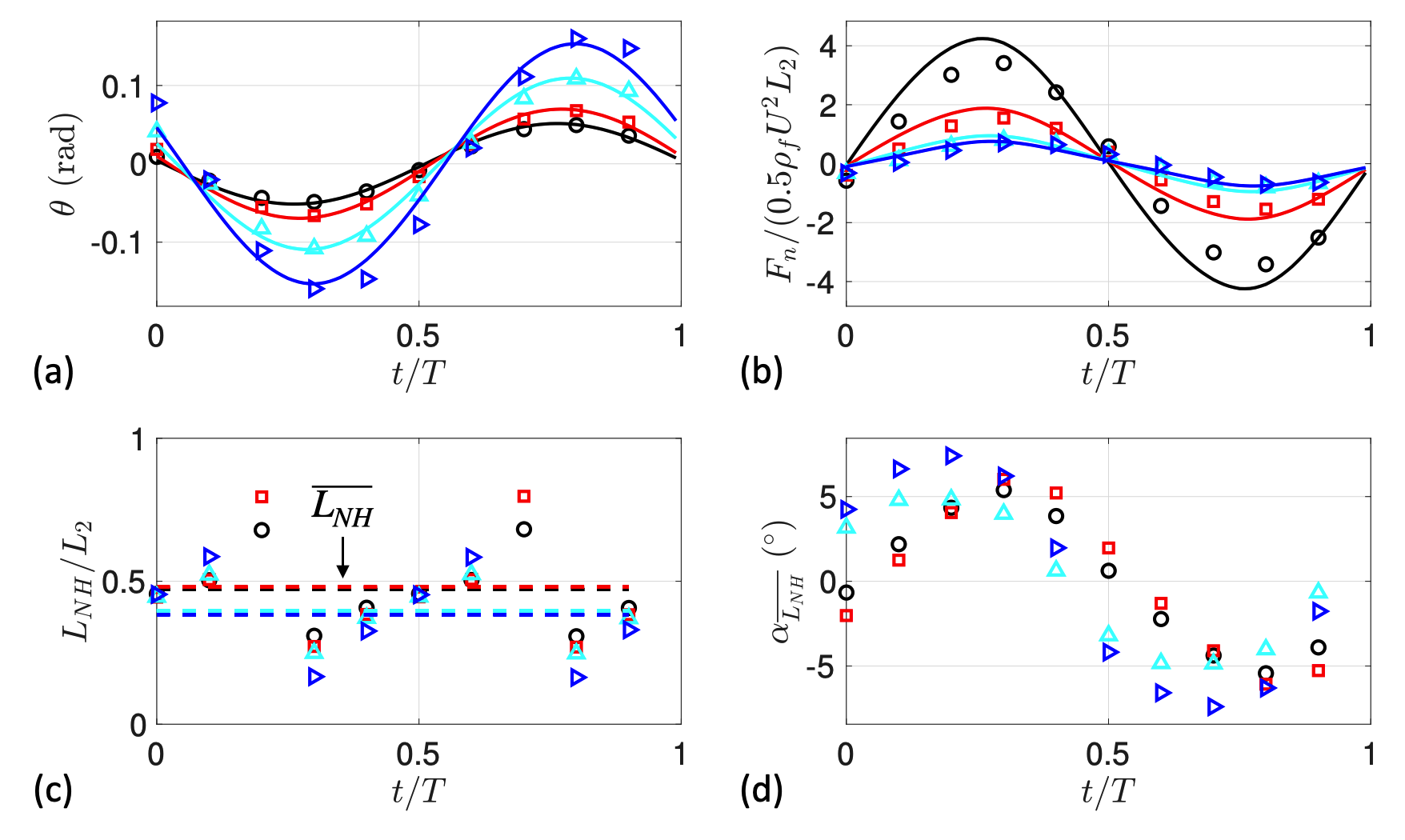}
\caption{(a-b) Comparison between Case 4 simulations (symbols) and model predictions ($\solid$), with different maximum flapping amplitude $a$: (a) $\theta$ and (b) $F_n$.
Comparison of (c) NH location and (d) angle of attack at the average NH location among simulation cases.
 $a=0.15$ ($\circ$), 0.2 ({\color{red}$\square$}) 0.3 ({\color{cyan}$\triangle$}) and 0.4 ({\color{blue}$\triangleright$}).
}
\label{fig:varA}	
\end{center}
\end{figure}

Figure~\ref{fig:varA} compares simulation and model results of Case 4, with various flapping amplitudes $a$. The variation of $a$ is found to affect noticeably the instantaneous and average NH locations, both moving downstream at a larger $a$ (Fig.~\ref{fig:varA}(c)). The local AOA values at the effective NH location are bounded between approximately $\pm 5$ degrees for all cases (Fig.~\ref{fig:varA}(d)).
The $\theta$ and $F_n$ values also depend sensitively on $a$ (Fig.~\ref{fig:varA}(a,b)). Such dependences are shown to be overall well predicted by the model.

Results in this section demonstrate that the model robustly predicts the FSI dynamics across a wide range of tail-beat parameters and Reynolds number. The small mismatch is probably because (i) the instantaneous NH location (i.e. that of zero AOA) is not fixed, but varies in time, and  (ii) the total fluid force is not exerted exactly at the NH location, but near the end of the body. 
Note that since the fluid drag used in the model is calibrated to yield the CFD-measured steady-swim velocity, the errors in $\theta$ and $F_n$ do not contain any contribution from an inaccurate swim velocity (which would require another fluid force model and would introduce additional error); instead, they are purely attributed to the slight mismatch in the NH location and the CP location as shown in Fig.~\ref{fig:aoa_dist_cases} for Case~1.

\section{Re$_L$-St relation }
\label{sec:re-st}

\begin{figure}  
\begin{center}
\includegraphics[width=1\linewidth]{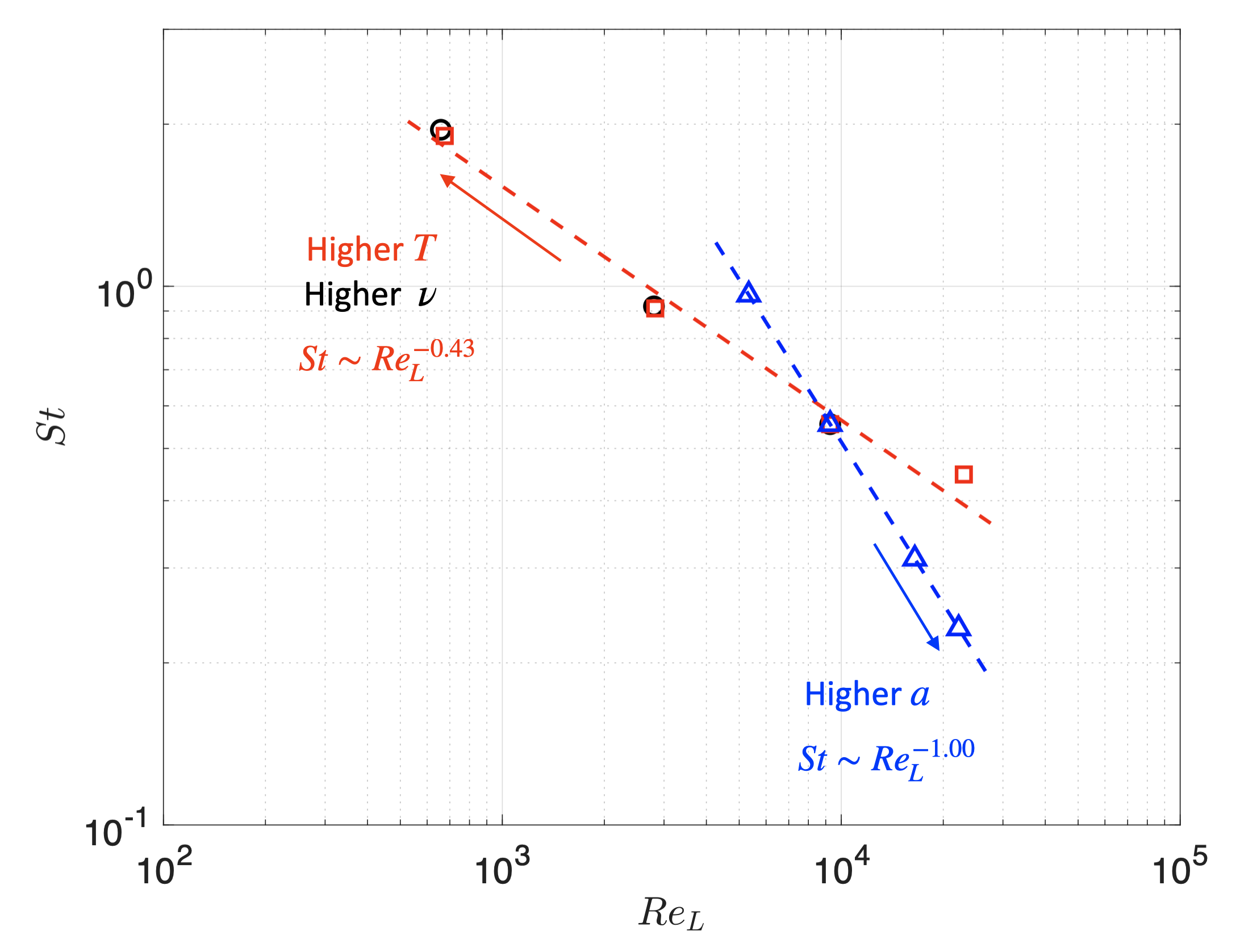}
\caption{The relation between Reynolds number and Strouhal number as shown by all simulation cases: Changing $\nu$  ($\circ$), changing $T$ ({\color{red}$\square$}) and changing $a$ ({\color{blue}$\triangle$}). $\dashed$ are fitted power-law relations. Arrows indicate direction of increase of parameter values.} 
\label{fig:re-st}	
\end{center}
\end{figure}

During steady swim, a one-to-one relationship exists between the Reynolds number (the dimensionless swim velocity) and the Strouhal number (the dimensionless tail-beat frequency). Based on a thrust-drag balance argument for fish swimming \cite{GazzolaAM14},  the scaling relation St$\sim$Re$_L^{-0.25}$ holds for the laminar flow regime, while St is approximately constant in the turbulent regime. This finding is supported by massive animal swim data from tadpoles to whales. These data also showed that the transitional Reynolds number (based on fish length) occurs at approximately $o(10^3)-o(10^4)$.  
Similar scaling relations were reported in previous numerical simulations of undulatory swim using actual fish geometries \cite[e.g.][]{MachucaGOB24,BorazjaniSotiropoulos08,BorazjaniBS09}. For example, using unsteady Reynolds averaged Navier-Stokes (URANS) simulations of swim of a lambari fish,  Mac{\'i}as et al. \cite{MachucaGOB24} observed a scaling relation of St$\sim$Re$_L^{-0.18}$, which differs from the relation reported by \cite{GazzolaAM14}  probably due to the different body geometry, gait and inherent uncertainties in the RANS turbulence closure.

Figure~\ref{fig:re-st} shows the scaling relation demonstrated based on the present data of the two-body system. The plot shows how varying different parameters (e.g., tail beat amplitude $a$, tail beat frequency $f$, and $\nu$) affects the scaling.
An inverse relation between Re$_L$ and St  is shown, regardless of which parameter is varied, consistent with previous observations. 
Interestingly, an increase of $f$ is found to be equivalent to a decrease of fluid viscosity in the parameter range simulated herein, shown by the two curves collapsing onto each other.
However, when $a$ varies,  St decreases with Re$_L$ faster than when $f$ or $\nu$ varies, as shown by the steeper slope of the fitted power-law relation. 
It is conjectured that, in the present parameter range, varying $f$ changes mostly the time scale of the problem, while varying $a$ changes the flow dynamics. This is supported by the observation of significantly different average NH locations observed when $a$ is varied (Fig.~\ref{fig:varA}(c)), compared to limited changes when $T$ or Reynolds number is varied (Fig.~\ref{fig:varT}(c) and Fig.~\ref{fig:Re_effect}(c)).
Although the nominal Re$_L$ values in the present data extend into the turbulent regime, the constant St scaling is not expected (as also shown in Fig.~\ref{fig:re-st}), as only the fin is resolved in this study and the flow is laminar around the fin.
The observed scaling provides evidence that the idealized two-body model accommodates some of the core dynamics of actual undulatory swim gait employed by swimming animals of all scales.
Note, however, that the scaling-relation comparison does not provide full validation of the low-order model, as currently the damping force in the low-order model is calibrated to match the steady-swim velocity in CFD (i.e. the Re-St relation is enforced to match the CFD data, see Sec.~\ref{sec:limitations}).

\begin{figure}[t]
\centering
\includegraphics[width=1\textwidth]{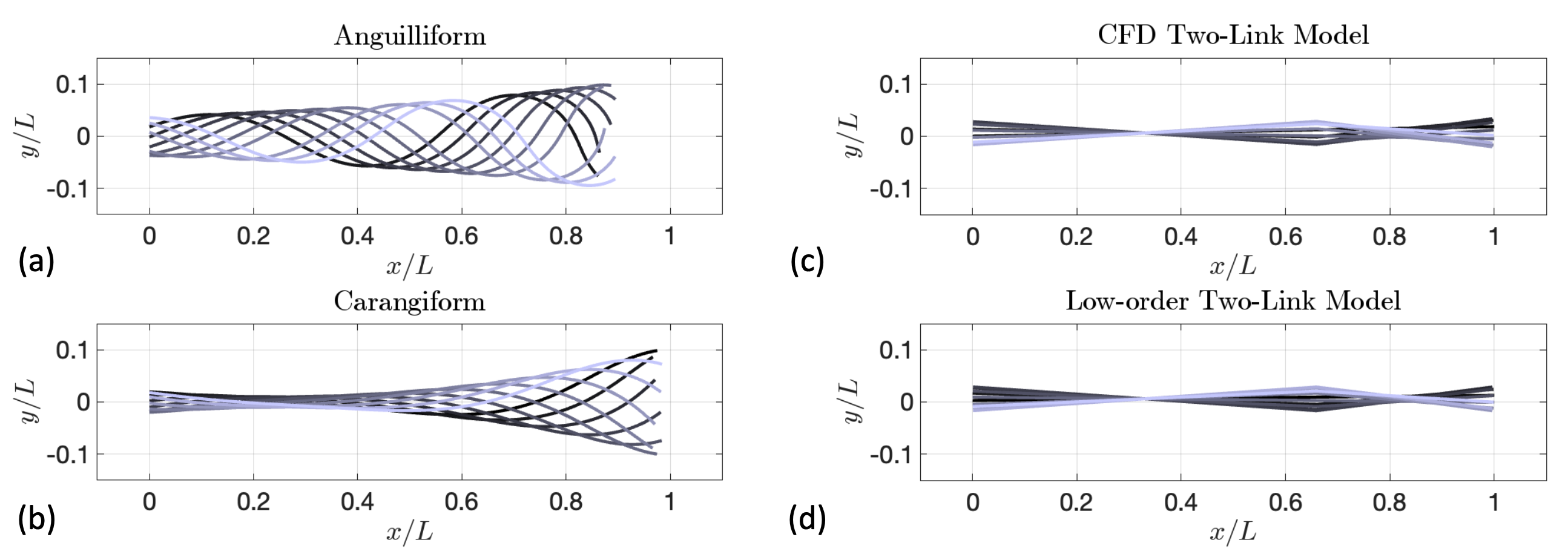}
\caption{ Comparison of actual fish swimming gaits: anguilliform (a, lamprey) and carangiform (b, mackerel), compared to the gaits of the two-link model from the present CFD (c) and low-order model (d). Note that (c) and (d) produce nearly identical gaits, consistent with the good body angle and force agreement shown earlier.}
\label{fig:gait_comparison}
\end{figure}

Comparing with actual swimming animals~\citep{GazzolaAM14} reveals quantitative differences: the present St values exceed the typical biological range of 0.2-0.4, though substantial variation exists in natural systems with St exceeding 1 at Re~$\sim$~O(10$^5$) in some cases. Note that the present Re is based on fin length while biological data typically use body length. The higher St values arise because the present Re values lie at the lower end of the biological spectrum, where higher Re generally corresponds to lower St until a transitional Re is reached. Additionally, kinematic differences exist between the two-link model and actual fish: in actual fish swimming (both anguilliform \cite{BorazjaniBS09} and carangiform \cite{BorazjaniSotiropoulos08} modes, Fig.~\ref{fig:gait_comparison}(a,b)), the deformation envelope increases continuously from head to tail, whereas the two-link model exhibits a different envelope due to the slot constraint (which forces the midpoint of body 1 to translate along a straight path)---the envelope is zero at the anchor point and varies non-monotonically along the body, leading to different force generation patterns at the fin location. Nevertheless, the carangiform mode shares some similarity with the two-link model during certain phases of the swimming cycle (Fig.~\ref{fig:gait_comparison}). Despite these quantitative differences, the model captures essential swimming dynamics: the wake eddy patterns in Fig.~\ref{fig:contours_slotcar}(c) demonstrate patterns characteristic of fish locomotion \cite{GazzolaAM14}.

\section{Discussion}
\label{sec:discussions}
\subsection{Key Insights}

The CFD simulations confirmed that while an exact fixed NH constraint location does not exist, an effective period-averaged NH location can be identified and applied. The variation in instantaneous NH location is expected given the unsteady flow and distributed fin forces, yet the period-averaged location provides a robust reference that captures essential dynamics. This suggests the NH constraint approach can extend to more complex swimming models using time-averaged constraint locations.

The observed Reynolds-Strouhal scaling (St$\sim$Re$_L^{-\alpha}$), consistent with biological data from tadpoles to whales \cite{GazzolaAM14}, demonstrates the two-body model captures fundamental hydrodynamic principles governing swimming across scales. The scaling behavior differs by parameter: varying tail-beat frequency produces effects equivalent to changing fluid viscosity (both curves collapse), while amplitude variations induce more complex flow changes reflected in altered NH constraint locations. This suggests frequency primarily affects time scale while amplitude alters flow dynamics—an insight for biomimetic design.

\subsection{Limitations}
\label{sec:limitations}

A key limitation is that the drag coefficient requires \textit{a priori} CFD calibration to match steady-swim velocity, preventing fully standalone predictions. Developing a self-contained drag model—potentially incorporating resistive force and pressure drag components \cite{Taylor1952, EloyE13}—is essential for a truly predictive low-order model and will be addressed in future work.

Additional limitations include the idealized two-rigid-body formulation, slot constraint restricting head motion to a straight line, and 2D nature of simulations. The 2D simulations neglect three-dimensional effects such as tip vortices forming at the fin edges, which influence thrust generation in actual swimming. The point-fin approximation, while validated for cases studied, may not capture distributed fin force effects in more realistic geometries. The simplified drag model assumes laminar flow over the fin. The two-rigid-body choice reflects a deliberate validation strategy: establishing NH constraint viability in the simplest configuration before pursuing more complex formulations.
Future extensions to multi-link formulations (three or more bodies) and relaxing the slot constraint at the midpoint of body 1 \cite{ArdisterFeeny2022b,ArdisterFeeny2023} better approximates continuous body deformation and would likely improve the Re-St scaling match with biological data.

\subsection{Implications}

This work provides quantitative validation of the two-body NH constraint model against fully resolved fluid-structure interaction using CFD. The NH constraint approach can extend to more realistic multi-link formulations without slot constraints.
Unlike traditional low-order models with restrictive flow assumptions (Section~1), the NH constraint accommodates steady and unsteady flows across Reynolds numbers and large-amplitude motions as long as the constraint holds. This broader applicability, combined with computational efficiency (orders of magnitude faster than CFD), makes it attractive for parameter studies, optimization frameworks, and bio-inspired robotics applications.

\section{Conclusions}

We studied a two-rigid-body undulatory swimmer where fin-fluid interaction was modeled using a nonholonomic constraint. The aims were to (a) demonstrate swimming-like locomotion and (b) validate the approach through CFD simulations.

We derived equations of motion using Lagrangian mechanics. Harmonic balance analyses aligned closely with numerical simulations. The model generates swimming-like locomotion with characteristic thrust patterns including frequency-doubling effects (two thrust peaks per stroke) and brief instances of negative thrust. Constraint force, thrust, and speed increased with amplitude $a$ and frequency $\omega$. Head oscillation increased with $a$ but not $\omega$. The model captures the Reynolds-Strouhal scaling (St$\sim$Re$_L^{-\alpha}$) observed across biological swimmers from tadpoles to whales. A key limitation is that the drag coefficient requires \textit{a priori} CFD calibration to match steady-swim velocity, preventing standalone predictions. A self-contained formulation is essential for future work.

To validate the model, we conducted CFD simulations with variations in fin size and location to match the low-order model parameters. The validation showed excellent agreement for body orientation angle and normal forces across variations in amplitude, period, and Reynolds number. While no exact fixed NH constraint location exists, an effective period-averaged location can be identified and applied.

This work provides quantitative validation of the nonholonomic constraint approach against fully resolved fluid-structure interaction, demonstrating broader applicability that avoids restrictive flow assumptions of resistive and reactive force models. Future work will evaluate the model with 3D CFD simulations, compare against existing force models, and extend to multi-link geometries and continuous body formulations without the slot constraint.

\vspace{1 cm}
\appendix

\section{Effect of Reynolds number}
\label{sec:A-re}

\begin{figure}[htbp] 
 \begin{center}
  \includegraphics[width=1\linewidth]{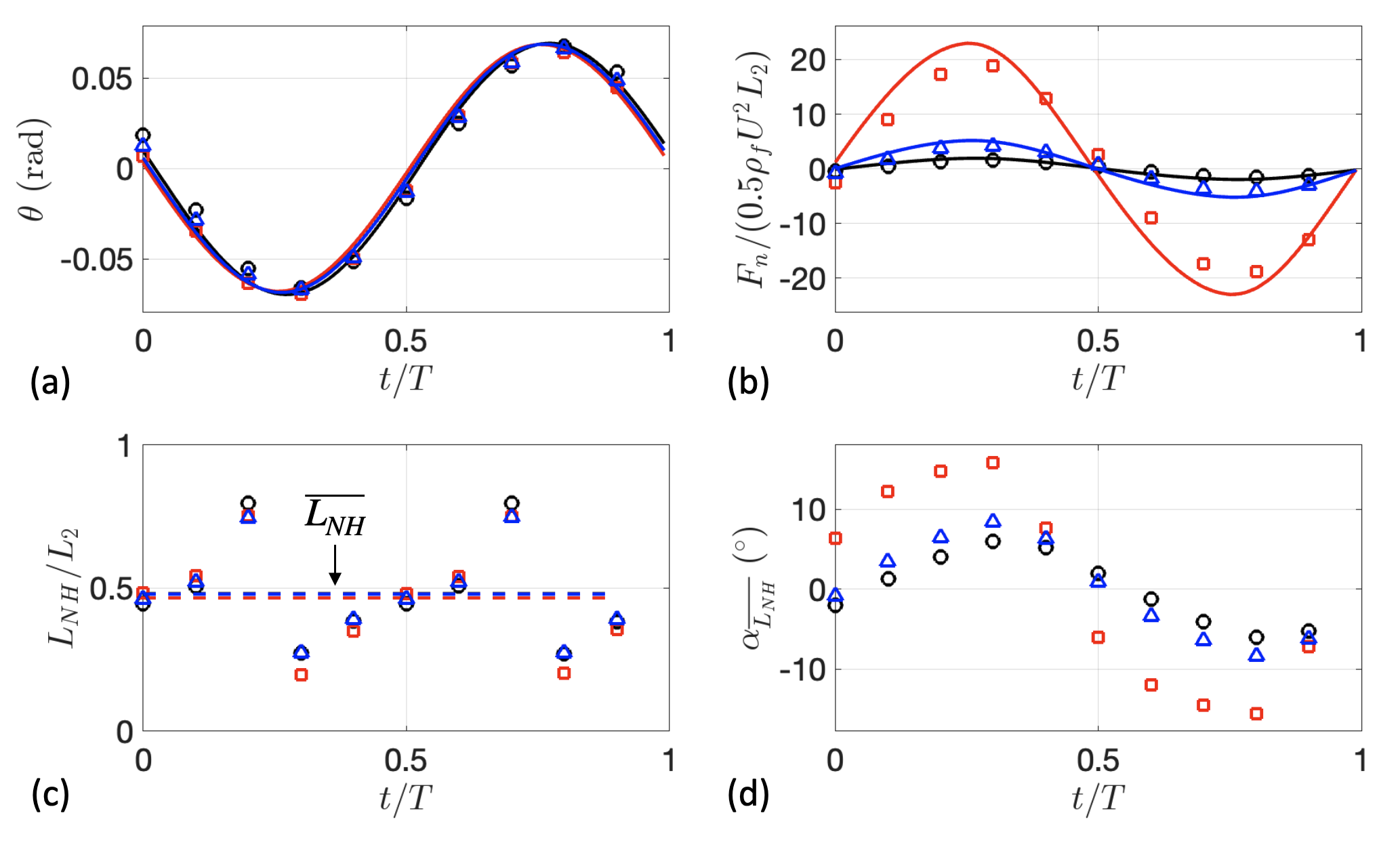}
  \caption{(a-b) Comparison between Case 4 simulations (symbols) and model predictions ($\solid$), with different Reynolds numbers: (a) $\theta$ and (b) $F_n$.
Comparison of (c) NH location and (d) angle of attack at the average NH location among simulation cases. 
Re$_L$ values are around 660 ({\color{red}$\square$}), 2800 ({\color{blue}$\triangle$}) and 9280 ($\circ$), respectively.
}
  \label{fig:Re_effect}	
 \end{center}
\end{figure}

The effect of varying the Reynolds number on the fluid-structure interaction is analyzed here. The three cases with different Reynolds numbers are achieved by progressively varying the fluid viscosity fourfold, while keeping the flapping period and amplitude constant. 
Figures~\ref{fig:Re_effect}(a,b) show that $\theta$ is insensitive to the Reynolds number, while the dimensionless normal force decreases with a higher Re. As $U$ increases with a higher Re (due to a lower viscous drag), this means that the increase of $F_n$ is not as fast as that of $U$ with an increase in Re.
The fin movement is rather insensitive to a change of Re: all Re values yield a similar period-average NH location (Fig.~\ref{fig:Re_effect}(c)), although the local AOA at the effective NH location appears to be independent of Reynolds numbers only at the two highest Reynolds number values (Fig.~\ref{fig:Re_effect}(d)). 
The AOA remains small at the mean NH location, maintaining approximate support of the NH model.

\section*{Acknowledgments}
This work was supported by the National Science Foundation, grant number 2015194.  Any findings, opinions, conclusions, or recommendations are those of the authors and do not necessarily reflect the views of the NSF.  

\bibliographystyle{elsarticle-num}
\bibliography{bibliography_consolidated}

\end{document}